\documentclass[prd,superscriptaddress,twocolumn,showpacs,showkeys,amsmath,%
preprintnumbers]{revtex4}
\usepackage{eucal} 
\usepackage{bm} 
\usepackage{graphicx}
\newcommand{\beq}{\begin{equation}}
\newcommand{\eeq}{\end{equation}}
\newcommand{\be}{\begin{eqnarray}}
\newcommand{\ee}{\end{eqnarray}}
\begin{document}
\author{A.~Afanasev}
\affiliation{Department of Physics, Hampton University, 
Hampton, VA 23668, USA}
\affiliation{Theory Center, Jefferson Lab, Newport News, VA 23606, USA}
\author{M.~Strikman}
\affiliation{Department of Physics, Pennsylvania State University,
University Park, PA 16802, USA}
\author{C.~Weiss}
\affiliation{Theory Center, Jefferson Lab, Newport News, VA 23606, USA}
\title{Transverse target spin asymmetry in inclusive DIS 
with two--photon exchange}
\begin{abstract}
We study the transverse target spin dependence of the cross 
section for inclusive electron--nucleon scattering with unpolarized beam.
Such dependence is absent in the one--photon exchange approximation 
(Christ--Lee theorem) and arises only in higher orders of the
QED expansion, from the interference of one--photon and absorptive 
two--photon exchange amplitudes as well as from real photon 
emission (bremsstrahlung). We demonstrate that the transverse 
spin--dependent two--photon exchange cross 
section is free of QED infrared and collinear divergences. 
We argue that in DIS kinematics the transverse spin dependence should be
governed by a ``parton--like'' mechanism in which the two--photon exchange 
couples mainly to a single quark. We calculate the normal
spin asymmetry in an approximation where the dominant contribution arises 
from quark helicity flip due to interactions with non--perturbative 
vacuum fields (constituent quark picture) and is proportional to
the quark transversity distribution in the nucleon. Such helicity--flip
processes are not significantly Sudakov--suppressed if the infrared 
scale for gluon emission in the photon--quark subprocess is of the order 
of the chiral symmetry breaking scale, 
$\mu^2_{\text{chiral}} \gg \Lambda^2_{\textrm{QCD}}$.
We estimate the asymmetry in the kinematics of the planned 
Jefferson Lab Hall A experiment to be of the order $10^{-4}$, 
with different sign for proton and neutron. We also comment on the 
spin dependence in the limit of soft high--energy scattering.
\end{abstract}
\preprint{JLAB-THY-07-717}
\maketitle
\section{Introduction}
Transverse spin effects in deep--inelastic $eN/\mu N$ scattering (DIS) are 
presently a very active field of research, with many interesting 
developments in experiment and theory. Inclusive production with 
longitudinally polarized beams and transversely polarized targets measures
the spin structure function $g_2$, which provides access to matrix
elements of higher--twist operators describing quark--gluon correlations 
in the nucleon \cite{Chen:2005td}. Another class of experiments measures 
the azimuthal distributions of identified hadrons in semi-inclusive 
production. The theoretical description of these observables relies on 
certain extensions of the usual collinear QCD expansion, which incorporate 
quark/hadron transverse momenta and give rise to a rich variety of 
distribution/fragmentation functions describing spin--orbit interactions 
of quarks. In analyzing the data one hopes to either learn about the 
spin--orbit interactions themselves or to use them to extract 
the quark transversity distributions in the nucleon.

A somewhat different transverse spin effect is the transverse target 
spin dependence of the cross section of inclusive DIS with unpolarized beam. 
Such dependence is absent in the $O(\alpha^2 )$ cross section in the
one--photon exchange approximation, being forbidden by the combination 
of $P$ and $T$ invariance and the hermiticity of the electromagnetic 
current operator (Christ--Lee theorem) \cite{ChL}. A target spin dependence 
appears at $O(\alpha^3)$ due to the interference of two--photon and 
one--photon exchange amplitudes, which can be understood qualitatively as 
the result of a non--hermitean effective current induced by the 
imaginary part of the two--photon exchange amplitude. 
Similar two--photon exchange effects were studied as corrections 
to the $eN$ elastic scattering cross section 
\cite{Blunden:2003sp,Guichon:2003qm,Chen:2004tw,Afanasev:2005mp},
where they partly explain the discrepancy between the $G_E/G_M$ ratio 
extracted using the Rosenbluth and polarization 
transfer methods \cite{Jones:1999rz,Gayou:2001qd}; they also play
a role in parity--violating electron scattering \cite{Afanasev:2005ex}.

The precision reached in $eN$ scattering experiments with modern 
high--duty cycle accelerators allows one to contemplate accurate 
measurements of two--photon exchange observables. A Jefferson Lab Hall A 
experiment \cite{PR-07-013} plans to measure the transverse target spin 
asymmetry in inclusive DIS ($E_{\text{beam}} = 6 \, \text{GeV},
x = 0.1 - 0.45, Q^2 = 1 - 3.5 \, \text{GeV}^2$) 
at the level of few times $10^{-4}$,
improving the sensitivity of the only previous measurement at
SLAC \cite{Rock:1970sj} by two orders of magnitude (in the SLAC
experiment, the asymmetry was found to be compatible with zero
at the level of $\sim 3.5\%$). It is timely to estimate the 
expected asymmetry in this kinematics.

In this paper we study the transverse target spin dependence in
inclusive DIS with unpolarized beam and its relation to 
the quark structure of the nucleon. This is a challenging problem,
combining the complexity of higher--order QED radiative corrections 
with that of the QCD treatment of transverse spin--dependent 
deep--inelastic processes. We approach this problem in steps,
establishing first some important general properties of the 
spin--dependent two--photon exchange cross section, then formulating
a scheme of approximations which respects these general properties
and allows us to estimate the expected asymmetry in DIS kinematics. 
In studying the general properties of the spin--dependent cross section
we shall employ both general principles [such as factorization of 
infrared (or IR) divergences, electromagnetic gauge invariance] 
and specific dynamical models which illustrate certain points.

First, we demonstrate that the transverse spin--dependent two--photon
exchange cross section is free of QED IR divergences. On general
grounds, it can be shown that the IR divergent terms take the form 
of a universal spin--independent factor multiplying the one--photon exchange 
cross section, which does not exhibit a transverse spin dependence. 
The IR finiteness can also be seen in the explicit expression 
for the two--photon exchange cross section for scattering from 
a spin--1/2 point particle. Furthermore, we show that the spin--dependent 
two--photon exchange cross section is free of QED collinear divergences,
which appear in intermediate stages of the calculation for 
a composite target with off--shell constituents. Such singularities
cancel as a consequence of electromagnetic gauge invariance. 
We illustrate this explicitly in a field--theoretical toy model of 
electron scattering from a spin--1/2 point particle dressed by a scalar field.

Second, we argue that the transverse spin--dependent cross section in 
DIS kinematics can be described in a ``parton--like'' picture, in which 
the two--photon exchange couples predominantly to a single quark,
namely the same quark which is hit in the interfering one--photon 
exchange process. Within this picture one then is dealing with two 
distinct contributions. In one the active quark helicity is conserved, 
but explicit interactions with the spectator system are required 
to bring about a non--zero result; this contribution is analogous 
to the twist--3 part of the spin structure function $g_2$. 
In the other, the quark helicity is flipped by interaction
with non--perturbative vacuum fields (spontaneous breaking of chiral symmetry)
and no interactions with the spectators are required; this contribution
is proportional to the product of the quark transversity distribution 
in the nucleon and the amplitude for the quark helicity--flip, which is
of the order of a typical ``constituent quark'' mass. In DIS such processes, 
going through low--virtuality quarks whose chirality 
can be flipped by vacuum fields, are in principle Sudakov--suppressed 
relative to those involving virtualities of the order $Q^2$. We show here
that the Sudakov suppression is not very effective if the IR
scale for gluon emission is of the order of the chiral symmetry 
breaking scale, $\mu^2_{\text{chiral}} \gg \Lambda^2_{\textrm{QCD}}$,
which appears natural from the point of view of the phenomenology
of spontaneous chiral symmetry breaking in QCD. A specific realization of
this scenario is the instanton vacuum model, in which the
chiral symmetry breaking scale is given by the average instanton size 
$\mu^2_{\text{chiral}} \sim \rho^{-2} \approx (600 \, \textrm{MeV})^2$
\cite{Diakonov:1985eg,Diakonov:2002fq}.

Third, in order to make a quantitative estimate we invoke the
additional approximation of a ``composite'' nucleon, \textit{i.e.},
a weakly bound system of constituent quarks, in which the dominant 
contribution to the spin--dependent cross section comes from helicity 
flip at the quark level and can be calculated in terms of the quark 
transversity distribution in the nucleon. This approximation permits 
a fully self--consistent treatment of the two--photon exchange cross section, 
which maintains electromagnetic gauge invariance and is manifestly free 
of collinear divergences. It allows us to make a numerical estimate of 
the asymmetry in DIS kinematics and discuss its dependence on the
kinematic variables.

We also comment on the behavior of the spin--dependent interference
cross section in the limit of soft high--energy scattering
(small energy and momentum transfer to the target), where one can 
make contact with general theorems about the high--energy behavior 
of QED amplitudes. Also, in this limit one can use the non--relativistic 
approximation to describe the target excitation spectrum and see 
explicitly why scattering from a single quark dominates at larger 
momentum transfers. This provides a useful complement to the 
corresponding arguments fielded in DIS kinematics.

Two--photon exchange effects were extensively studied as corrections 
to the $eN$ elastic scattering cross section 
\cite{Blunden:2003sp,Guichon:2003qm,Chen:2004tw,Afanasev:2005mp}.
The two--photon exchange effect in the transverse spin dependence
investigated in this article is in many ways simpler than those corrections
to the cross section. In the transverse spin--dependent cross section 
one is dealing with a pure higher--order QED observable, which is exactly
zero in one--photon exchange approximation. More importantly,
because the two--photon exchange in the transverse spin--dependent 
cross section is IR--finite, no cancellations of IR
divergences between two--photon exchange and real photon bremsstrahlung
take place as in the spin--independent cross section. In fact, this 
circumstance makes it possible to discuss two--photon exchange as an
``autonomous'' physical effect in the first place.

This article is organized as follows. In Section~\ref{sec:asymmetry}
we define the transverse spin--dependent cross section and review the 
Christ--Lee theorem for the one--photon exchange contribution. 
In Section~\ref{sec:point} we revisit in some
detail the transverse spin--dependent cross section in the scattering 
from a pointlike spin--1/2 particle, which explicitly shows the IR 
finiteness and allows us to estimate the effective photon virtualities 
in the two--photon exchange. In Sec.~\ref{sec:divergences} we demonstrate 
the absence of QED IR and collinear divergences in the transverse 
spin--dependent cross section on general grounds. In Sec.~\ref{sec:qcd} we 
consider the transverse spin dependence in DIS in QCD. We present arguments 
in favor of dominance of scattering from a single quark, discuss the 
two contributions (quark helicity--conserving and quark helicity--flip), 
and the absence of significant Sudakov suppression of quark helicity--flip 
amplitudes. In Sec.~\ref{sec:composite} we formulate the composite nucleon
approximation, in which the quark helicity--flip contribution 
becomes dominant and can be calculated in a relativistic constituent 
quark model. In Sec.~\ref{sec:numerical} we present numerical results
based on this approximation. In Sec.~\ref{sec:soft} we discuss the
limit of soft high--energy scattering. Our conclusions and perspectives
for future studies are summarized in Sec.~\ref{sec:summary}.
%
%
\begin{figure*}
\begin{tabular}{ccc}
\includegraphics[width=5cm]{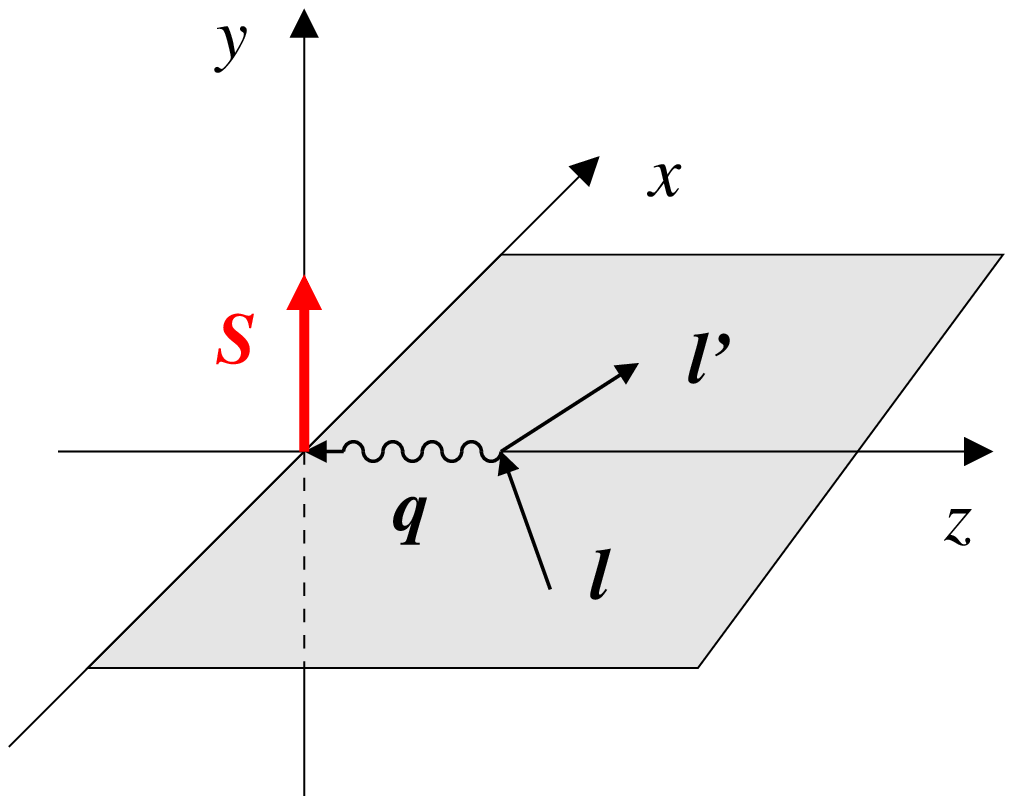}
& \hspace{5em} &
\includegraphics[width=5cm]{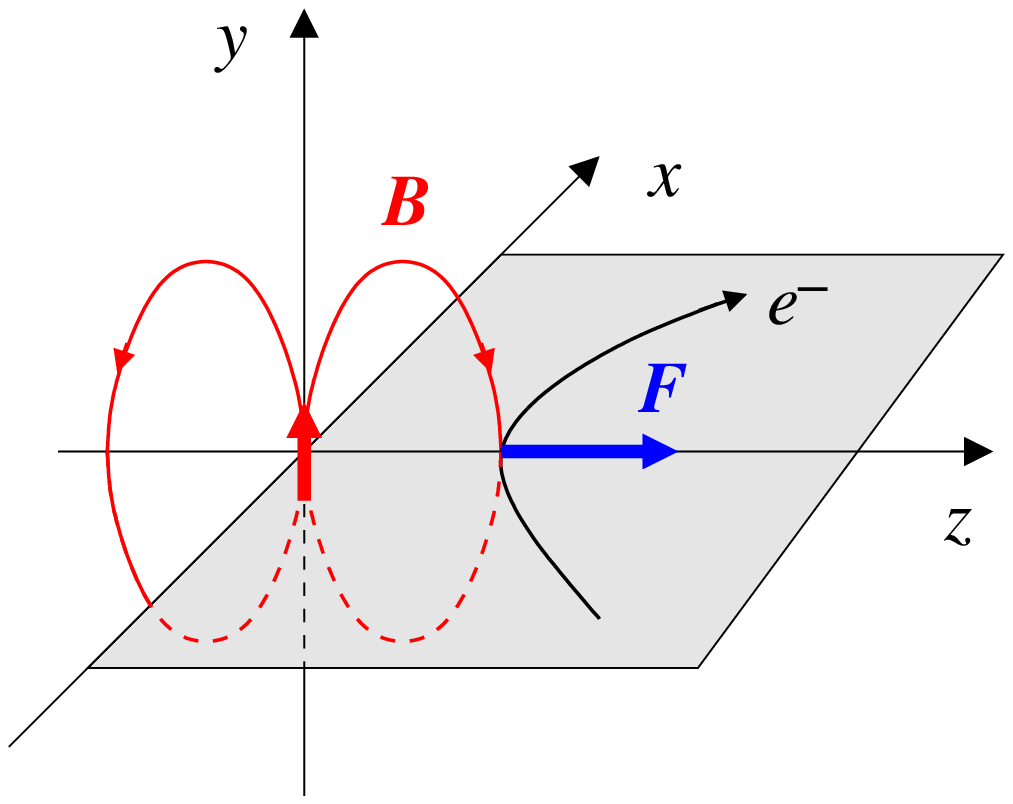}
\\
(a) & & (b)
\end{tabular}
\caption[]{(a) The coordinate system for describing the transverse
spin dependence of inclusive DIS in the target rest frame. 
(b)~Transverse target spin 
asymmetry in the scattering of an electron (charge $-e$) from a 
classical pointlike magnetic dipole. The asymmetry results from 
the Lorentz force, $\bm{F}$, experienced by the charge moving in
the magnetic field of the dipole, $\bm{B}$.}
\label{fig:xyz}
\end{figure*}
\section{Transverse target spin in inclusive electron scattering}
\label{sec:asymmetry}
We consider inclusive electron--nucleon scattering
with unpolarized beam and polarized target,
\beq
e(l) + N(p) \;\; \rightarrow \;\; e(l') + X .
\label{inelastic}
\eeq
For parity--conserving interactions (strong, electromagnetic) 
the only allowed dependence of the cross section in the target 
rest frame on the target spin is through a term proportional to
the true scalar
\beq
(\bm{S}, \bm{l} \times \bm{l}') .
\eeq
Here $\bm S$ denotes the target polarization vector, which is normalized 
according to $\bm{S}^2 = 1$ for fully polarized target, and the vector
$\bm{l} \times \bm{l}'$ is normal to the electron scattering plane. 
We write the differential cross section for scattering into a phase 
space element with given final electron momentum $\bm{l}'$ in the form 
\beq
d\sigma \;\; = \;\;
d\sigma_U  \; + \;  
\frac{(\bm{S}, \bm{l} \times \bm{l}')}{|\bm{l} \times \bm{l}'|}
\; d\sigma_N .
\label{dsigma}
\eeq
The normal spin asymmetry of the differential cross section is then 
defined as
\beq
A_N \;\; \equiv \;\; \frac{d\sigma_N}{d\sigma_U} .
\eeq
It can be interpreted as the asymmetry of the differential cross section 
for scattering to the ``left'' and ``right'' of a target polarized 
``upward'' in the direction normal to the scattering plane,
with otherwise identical kinematics,
\beq
A_N \;\; = \;\; \frac{d\sigma (\text{left}) - d\sigma (\text{right})}
{d\sigma (\text{left}) + d\sigma (\text{right})} .
\label{A_left_right}
\eeq

To describe transverse spin effects in DIS kinematics it is 
customary to define a coordinate system such that the 
momentum transfer 
\beq
\bm{q} \equiv \bm{l} - \bm{l}' 
\label{q}
\eeq
(\textit{i.e.}, the momentum of the virtual photon in one--photon 
exchange approximation) points in the negative $z$--direction,
and the initial and final electron momenta lie in the $xz$--plane, 
with the average momentum pointing in the positive $x$--direction
(see Fig.~\ref{fig:xyz}a). In this frame the unit vector 
$\bm{l} \times \bm{l}' / |\bm{l} \times \bm{l}'|$
points in the negative $y$--direction, and the normal
spin asymmetry coincides with the negative polarization
asymmetry with respect to the target spin in the $y$--direction,
\beq
A_N \;\; = \;\; \frac{d\sigma (S_y = -1) - d\sigma (S_y = +1)}
{d\sigma (S_y = -1) + d\sigma (S_y = +1)} \;\; \equiv \;\; -A_y.
\label{A_y}
\eeq
It is clear that this definition applies not only to the target
rest frame but also to the virtual photon--nucleon center--of--mass 
(CM) frame, in which the nucleon moves in the positive $z$ direction.

The cross section for inclusive $eN$ scattering with unpolarized 
beam is independent of the transverse target spin if the 
electromagnetic interaction is treated in one--photon 
exchange approximation (Christ--Lee theorem) \cite{ChL}. 
In this approximation the cross section can be expressed in 
the well--known form \cite{LLIV}
\beq
d\sigma \;\; = \;\;
\frac{e^4}{4 (lp) \, Q^4}
\; L^{\mu\nu} \, W_{\mu\nu} \; \frac{d^3 l'}{(2\pi)^3 \, 2E'} ,
\label{cross_section_from_tensors}
\eeq
where $e$ is the elementary charge and
\be
Q^2 &\equiv& -q^2 \;\; = \;\; - (l' - l)^2
\ee
the invariant momentum transfer. The leptonic tensor, 
$L_{\mu\nu}$, is symmetric for an unpolarized beam, 
$L_{\mu\nu} = L_{\nu\mu}$, and the
contraction in Eq.~(\ref{cross_section_from_tensors}) 
projects out the symmetric part of the hadronic tensor,
\beq
W_{\mu\nu} \;\; = \;\; \int d^4x \; e^{i (qx)}
\; \langle p S | J_\mu (x) J_\nu (0) | p S \rangle .
\label{W_mu_nu_def}
\eeq
Using $P$ and $T$ invariance as well as the hermiticity of the 
current operator, it can be shown that the symmetric part of
the hadronic tensor remains unchanged under reversal of the
target's transverse polarization, and the asymmetry (\ref{A_y}) 
is zero. We shall see an explicit example of this general theorem
in the cross section for a pointlike target in Sec.~\ref{sec:point}.

\begin{figure}
\includegraphics[width=0.4\textwidth]{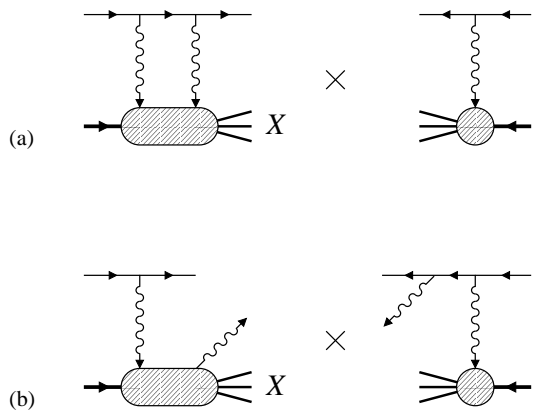}
\caption[]{QED processes contributing to the transverse
target spin dependence of the inclusive $eN$ cross section at $O(\alpha^3 )$.
(a)~Interference of one--photon and two--photon exchange.
(b)~Interference of real photon emission (bremsstrahlung) by 
the electron and the hadronic system.}
\label{fig:twogamma}
\end{figure}

Target spin dependence in $P$-- and $T$--invariant inclusive scattering 
can arise only from higher--order electromagnetic interactions.
At the order $\alpha^3$, one can identify two distinct contributions
to the cross section which give rise to a target spin dependence, see
 Fig.~\ref{fig:twogamma}. One is the interference of one--photon 
and two--photon exchange amplitudes in the $ep \rightarrow e'X$ 
cross section (Fig.~\ref{fig:twogamma}a). This mechanism can,
in a sense, be regarded as a non--hermitean contribution to the 
current operator in the leading--order expression, arising from the 
imaginary part of the two--photon exchange contribution to the 
$ep \rightarrow e'X$ amplitude. The other contribution results from the 
interference of real photon radiation (bremsstrahlung) emitted by the 
electron and the interacting hadronic system (Fig.~\ref{fig:twogamma}b). 
An important point is that the two--photon exchange contribution to the 
spin--dependent cross section, Fig.~\ref{fig:twogamma}a, is free of QED 
IR divergences, as will be discussed in detail in Sec.~\ref{sec:divergences}. 
In the cross section spin difference and the asymmetry (\ref{A_y}) 
two--photon exchange and real photon emission can thus be regarded as 
physically distinct contributions from the QED point of view and 
be discussed separately. This is in contrast to the two--photon exchange 
contributions to the cross section itself for given target spin 
(or the sum over target spins), where the IR divergences cancel only 
when two--photon exchange and real photon emission are added.
%
%

A non--zero target spin dependence of inclusive $eN$ scattering 
could in principle arise if $T$ invariance were violated explicitly in 
electroweak interactions. In fact, the SLAC experiment \cite{Rock:1970sj}
measured the spin asymmetry with the aim of testing $T$ invariance 
of the $ep$ interaction and found the asymmetry be consistent 
with zero at the level of 3.5\%. Present understanding of 
the limits on the violation of fundamental symmetries suggests that
$P$--conserving, $T$--violating effects in the Standard Model,
which come as weak interaction corrections to $P$--violating effects, should
lead to corrections to the DIS cross section of the order of 
at most $< 10^{-8}$ \cite{Conti:1992xn}. These effects are significantly 
smaller than the asymmetry expected from two--photon exchange, 
$|A_N| \sim 10^{-4}$ (see below). The $T$--violating effects could
in principle be separated from two--photon exchange by their different 
beam charge dependence \cite{Rock:1970sj}. However, electromagnetic 
effects at $O(\alpha^4 )$, such as three--photon exchange and
double two--photon exchange, would have the same spin and beam charge 
dependence as $T$--violation and exceed the latter by at least two orders 
of magnitude, making it practically impossible to probe explicit
$T$--violation in this way.

\section{Transverse spin dependence for pointlike target}
\label{sec:point}
We begin our investigation of the transverse spin dependence
by considering the scattering of an electron (charge $-e$) from a 
Dirac point particle of charge $+e$, referred to as ``pointlike proton'' 
in the following. While several calculations of the asymmetry in 
this model have been reported long ago \cite{Barut60,DeRujula:1972te}, 
it is worthwhile to revisit this problem for several reasons. 
First, the point particle calculation explicitly demonstrates the 
IR--finiteness of the two--photon exchange contribution to the asymmetry, 
and allows us to investigate numerically the distribution of photon 
virtualities in the two--photon exchange graph. Second, the point particle 
result provides a crude --- but manifestly self--consistent --- estimate of 
the asymmetry, including real photon emission (which turns out not
to contribute to the asymmetry in this case), and will serve as a 
reference point for more elaborate models including hadron structure. 
Third, we need the point particle result as an ingredient for the 
composite nucleon approximation in Sec.~\ref{sec:composite}.

\begin{figure}
\includegraphics[width=8.5cm]{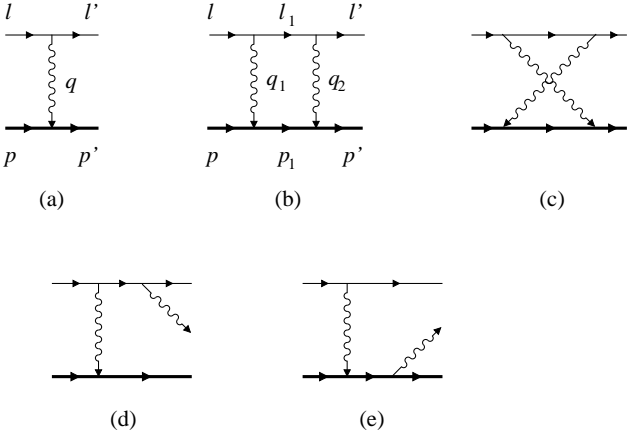}
\caption[]{Feynman diagrams contributing to the amplitude of
$ep$ scattering to order $\alpha^2$. (a, b, c) Elastic scattering,
$ep \rightarrow e'p$. The two--photon box diagram (b) gives rise to 
an imaginary part of the amplitude. (d, e) Real photon emission.}
\label{fig:point}
\end{figure}

For a pointlike proton, the hadronic final state in inclusive $ep$ 
scattering contains just the proton itself. Likewise,
there are no excited hadronic intermediate states in higher--order
processes. The Feynman diagrams contributing to the amplitude
for inclusive $ep$ scattering to order $\alpha^2$ are shown 
in Fig.~\ref{fig:point}. Consider first the elastic scattering
channel,
\beq
e(l) + p(p) \;\; \rightarrow \;\; e(l') + p(p') ,
\label{elastic}
\eeq
the amplitude of which is given by the sum of diagrams (a)--(c).
%
%
In this channel the sum over hadronic final states reduced to
the sum over the final--state proton polarization states. 
The cross section for scattering from a transversely polarized proton
is proportional to the squared modulus of the invariant amplitude, 
averaged (summed) over the initial (final) electron polarization, 
and summed over the final proton polarization. On general grounds, 
the dependence of this quantity on the polarization of the initial proton
must be of the form
\beq
|\mathcal{M}_{ep \rightarrow e'p}|^2 \;\; = \;\; X_U
\; - \; \frac{(SN)}{\sqrt{-N^2}} \; X_N ,
\label{M2_from_X}
\eeq
where $S$ is the polarization 4--vector of the 
initial proton state, 
\beq
N^\mu \;\; = \;\; 
-4 \epsilon^{\mu\alpha\beta\gamma} l_\alpha l'_\beta p_\gamma 
\label{N}
\eeq
the normal 4--vector characterizing the scattering process 
\cite{foot:conventions}, and $X_U$ and $X_N$ are independent 
of the initial proton polarization. Noting that in the target rest 
frame $S^\mu = (0, \bm{S})$, and $N^\mu = (0, \bm{N})$ with 
$\bm{N} = 4 M \bm{l} \times \bm{l}'$ normal to the scattering plane, 
we have
\beq
- \, \frac{(SN)}{\sqrt{-N^2}} \;\; = \;\; 
\frac{(\bm{S}, \bm{l} \times \bm{l}')}{|\bm{l} \times \bm{l}'|} ,
\eeq
and the transverse target spin asymmetry (\ref{A_y}) is given by
\beq
A_N \;\; = \;\; \frac{X_N}{X_U} .
\label{A_from_X}
\eeq
This representation allows us to calculate the asymmetry directly 
from the invariant amplitudes, without reference to a particular frame. 
We note that in the frame of Fig.~\ref{fig:xyz}a, choosing the proton
polarization to be along the $y$ axis, the coefficients in 
Eq.~(\ref{M2_from_X}) are given by
\be
X_U &=& \frac{1}{2} \left[ \, |\mathcal{M}(y-)|^2 + |\mathcal{M} (y+)|^2 
\, \right] ,
\\
X_N &=& \frac{1}{2} \left[ \, |\mathcal{M}(y-)|^2 - |\mathcal{M} (y+)|^2 
\, \right] ,
\ee
where $\mathcal{M} (y\pm ) \equiv \mathcal{M}_{ep \rightarrow e'p} (y\pm )$ 
denotes the amplitude for scattering from a proton
state with $S_y = \pm 1$. In this case expression (\ref{A_from_X})
for the asymmetry reproduces the negative $y$ spin asymmetry,
Eq.~(\ref{A_y}). It is instructive to express the coefficients
$X_U$ and $X_N$ also in terms of the amplitudes for scattering
from a proton of given helicity. In a frame where the proton
moves in the positive $z$ direction the helicity eigenstates 
$|\pm\rangle$ coincide with the eigenstates of $S_z$ and are 
related to the $S_y$ eigenstates by
\beq
|y\pm\rangle \;\; = \;\; 
\frac{|+\rangle \pm i |-\rangle}{\sqrt{2}} ,
\eeq
and one obtains
\be
X_U &=& \frac{1}{2}
\left[ \, |\mathcal{M}(-)|^2 +  |\mathcal{M} (+)|^2 \, \right] ,
\\[1ex]
X_N &=& \text{Im} \left[ \mathcal{M}^\ast (-)  \mathcal{M} (+) \right] .
\label{X_N_from_helicity}
\ee
In the helicity basis the transverse spin dependence is related to the 
interference of helicity--flip and non--flip amplitudes in the cross section. 
In particular, it is seen from Eq.~(\ref{X_N_from_helicity}) that a 
spin dependence appears only if the helicity amplitudes develop 
an imaginary (absorptive) part.

In the approximation of zero electron mass, $m \rightarrow 0$,
the electron helicity is conserved because of chiral invariance, 
and $ep$ elastic scattering (\ref{elastic}) is described by 3 independent 
helicity amplitudes. We parametrize the invariant amplitude as
\be
\mathcal{M}_{ep \rightarrow e'p} &=& \bar u' \hat P u \; \left( 
2 M  \bar U' U \; f_1 \; + \; \bar U' \hat L U \; f_2 \right)
\nonumber \\
&+& \bar u' \hat P \gamma_5 u \; \bar U' \hat L \gamma_5 U \; f_3 ,
\label{M_thru_f}
\ee
where $u, u'$ and $U, U'$ are the bispinors of the initial/final 
electron and proton, normalized as $\bar u u = 2 m, \bar U U = 2 M$, 
$M$ is the proton mass, 
\be
L &\equiv& l + l' , 
\label{L}
\\
P &\equiv& p + p' 
\ee
are the sum of the initial and final electron/proton momenta,
and we use the notation $\hat P \equiv P^\mu \gamma_\mu$.
Here $f_1$--$f_3$ are scalar functions of the kinematic invariants, 
\be
s &\equiv& (l + p)^2 , \\
t &\equiv & (l' - l)^2 \;\; = \;\; q^2;
\ee
it is convenient to introduce also the crossing--symmetric variable
\beq
\nu  \;\; \equiv \;\; (LP) \;\; = s - u \;\; = \;\;
2 (s - M^2) + t . 
\eeq
By straightforward calculation, using the standard expressions for 
the spin density matrices of the electron and proton spinors
\cite{foot:conventions}, one obtains the coefficients of the 
squared modulus of the invariant amplitude, Eq.~(\ref{M2_from_X}), as
\be
X_U &=& \left[ \nu ^2 - t (t - 4 M^2) \right]
\nonumber \\
&\times & \left\{ 4 M^2 (-t + 4 M^2) \; |f_1|^2  \right.
\nonumber \\
&+& ( \nu^2 - t^2 )\; |f_2|^2 
\nonumber \\
&+& 8 M^2 \nu \; \text{Re}(f_1^\ast f_2) 
\nonumber \\
&+& \left. \left[ \nu ^2 - t (t - 4 M^2) \right] \;  |f_3|^2 \right\} ,
\\
X_N &=& 4 M \; \left[ \nu ^2 - t (t - 4 M^2) \right]
\nonumber \\
&\times& \sqrt{-t \left[ \nu^2 - t (t - 4 M^2) \right]} 
\; \text{Im} (f_1^\ast f_2) .
\label{X_N_from_f}
\ee
Again, one sees from Eq.~(\ref{X_N_from_f}) that a spin dependence 
of the cross section appears only if the functions
$f_1$ and $f_2$ develop an imaginary part. 

In one--photon exchange approximation the invariant amplitude
for elastic $ep$ scattering is given by the diagram of 
Fig.~\ref{fig:point}a,
\beq
\mathcal{M}_{ep \rightarrow e'p}^{\text{(a)}} \;\; = \;\; -\frac{e^2}{t} \;
\bar u' \gamma^\mu u \; \bar U' \gamma_\mu U ,
\eeq 
where the negative sign results from the different sign of the charges.
The contribution to the functions $f_1$--$f_3$ can easily be
found by expanding the vector currents in the basis formed
by the orthogonal 4--vectors $L, q, N$ and $P - (LP) L / L^2$,
and using the relations between bilinear forms following from the 
three--gamma identities and the Dirac equation for the electron 
and proton spinors \cite{LLIV}. One obtains
\beq
\left.
\begin{array}{r}
f_1^{\text{(a)}} \\[3ex]
f_2^{\text{(a)}} \\[3ex]
f_3^{\text{(a)}}
\end{array}
\right\}
\;\; = \;\; 
\frac{-e^2}{\nu^2 - t (t - 4M^2)} \; \times
\left\{
\begin{array}{r}
\displaystyle 1 , \\[3ex]
\displaystyle \frac{\nu}{t} , \\[3ex]
\displaystyle -1 .
\end{array}
\right.
\eeq
In this approximation the functions $f_1$--$f_3$ are real,
and the spin--dependent part of the squared invariant amplitude
(\ref{X_N_from_f}) is zero,
\beq
X_N^{\text{(a)}} \;\; = \;\; 0 ,
\eeq
in accordance with the Christ--Lee theorem. The spin--independent
part gives the usual expression for the squared amplitude in
unpolarized $ep$ elastic scattering,
\beq
X_U^{\text{(a)}} \;\; = \;\;
\frac{e^4}{t^2} \left[ \nu^2 + t (t + 4 M^2) \right] .
\label{X_U_a}
\eeq

%
%
\begin{figure*}
\begin{tabular}{ccc}
\includegraphics[width=7cm]{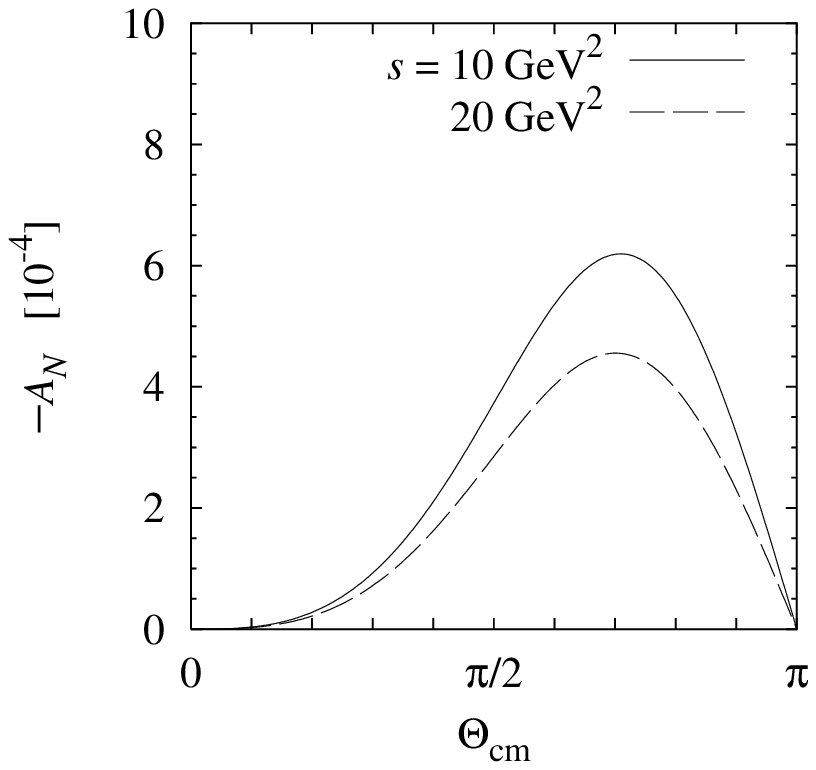}
&
\hspace{1cm}
&
\includegraphics[width=7cm]{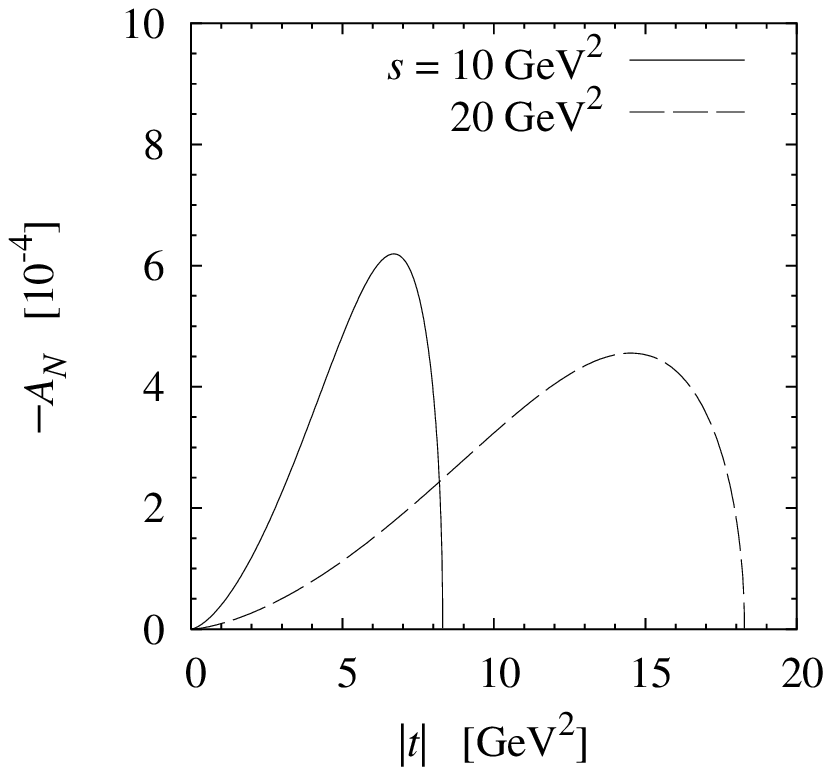}
\end{tabular}
\caption[]{The normal spin asymmetry $A_N$ (note the minus sign on the axis)
in electron scattering from a pointlike proton, for $s = 10$ and 
$20 \, \text{GeV}^2$, as a function of the CM scattering angle (left), 
and of $|t| = Q^2$ (right).}
\label{fig:an_tdep}
\end{figure*}
A non--zero imaginary part of $f_1$--$f_3$ arises 
at order $\alpha^2$ from the two--photon exchange box diagram, 
Fig.~\ref{fig:point}b. (The crossed--box diagram, Fig.~\ref{fig:point}c,
does not have an imaginary part in the physical region for $ep$
scattering.) The contribution of diagram Fig.~\ref{fig:point}b
to the invariant amplitude is given by the Feynman integral
\be
\lefteqn{
\mathcal{M}_{ep \rightarrow e'p}^{\text{(b)}} 
\;\; = \;\; -i \; \int \frac{d^4 \Delta}{(2\pi)^4}}
&& 
\nonumber \\
&\times& \frac{e^4 \; \bar u' \gamma^\mu \hat l_1 \gamma^\nu u
\;\; \bar U' \gamma_\mu (\hat p_1 + M) \gamma_\nu U}
{q_1^2 \; q_2^2 \; (l_1^2 + i0) \; (p_1^2 - M^2 + i0)} ,
\label{M_b_feynman}
\ee
where $\Delta$ represents a suitably chosen loop momentum, 
\textit{e.g.},
\be
q_{1,2} &=& \; q/2 \pm \Delta, \\
l_1 &=& L/2 - \Delta , \\
p_1 &=& P/2 + \Delta .
\ee
By projecting the numerator in Eq.~(\ref{M_b_feynman}) on the 
structures of Eq.~(\ref{M_thru_f}), using the basis vectors described
above, one can easily determine the corresponding contributions to 
the functions $f_1$--$f_3$.
Their imaginary part is then calculated by applying the Cutkosky rules, 
replacing the propagators of the intermediate particles by delta 
functions. We are interested only in the interference term, 
$\text{Im} \, (f_1^\ast f_2)$, which governs the cross section 
spin difference, Eq.~(\ref{X_N_from_f}). Because the 
one--photon exchange amplitudes are real, the leading non--zero 
contribution to this term is
\beq
\text{Im} \, (f_1^\ast f_2) \;\; = \;\; 
f_1^{\text{(a)}} \, \text{Im} \, f_2^{\text{(b)}} - 
f_2^{\text{(a)}} \, \text{Im} \, f_1^{\text{(b)}} .
\label{f_combination}
\eeq
It is convenient to combine the functions in this way before 
performing the loop integral. In this way one obtains a 
representation of the interference term as
\be
\text{Im} \, (f_1^\ast f_2) 
&=& 2 \pi^2  \; 
\int \frac{d^4 \Delta}{(2\pi)^4} \;
\delta (l_1^2 ) \; \delta (p_1^2 - M^2) \; \frac{\phi_A}
{q_1^2 \; q_2^2} , 
\nonumber \\
\label{Im_f1_f2_integral} 
\ee
where the integration is restricted over positive--energy 
intermediate states, $(l_1)^0, (p_1)^0 > 0$, 
and the numerator is given by
\be
\phi_A &=& \frac{-e^6}{\nu^2 - t (t - 4 M^2)}
\nonumber \\
&\times&
\left\{ \frac{1}{2t^2} \; (q^2 + q_1^2 - q_2^2)(q^2 - q_1^2 + q_2^2)
\right.
\nonumber \\
&+& \frac{1}{4t} \; \left[ (q_1 - q_2)^2 - q^2 \right]
\nonumber \\
&+& \frac{3 (\nu - t) + 4 M^2}{8 t [\nu^2 - t (t - 4 M^2)]}
\; \left[ (q_1 - q_2)^2 - q^2 \right] 
\nonumber \\
&& \left.\phantom{\frac{1}{1}} \times \left[ (q_1 - q_2)^2 + q^2 \right]
\right\} ,
\label{phi_A_explicit}
\ee
where $q^2 = t$. In simplifying Eq.~(\ref{phi_A_explicit}) we have made 
use of the mass shell conditions $l_1^2 = 0$ and $p_1^2 = M^2$
implied by the delta functions. Equations~(\ref{Im_f1_f2_integral}) 
and (\ref{phi_A_explicit}) allow us to evaluate the spin--dependent 
cross section as an invariant integral.

An important observation is that the numerator (\ref{phi_A_explicit})
vanishes at in the limits where the 4--momentum of one or the other 
photon in the two--photon exchange graph vanishes,
\beq
\phi_A \;\; \rightarrow \;\; 0 
\;\;\;\; \text{for} \;\;\;\; 
\left\{
\begin{array}{ll} 
q_1 \; \rightarrow \; 0, \;\; & q_2 \; \rightarrow \; q, \;\; \text{or}
\\
q_2 \; \rightarrow \; 0, & q_1 \; \rightarrow \; q .
\end{array}
\right.
\eeq
This implies that the integral representing the spin--dependent
interference cross section (\ref{Im_f1_f2_integral}) is free of
IR divergences. We shall see in Sec.~\ref{sec:divergences}
that this property is general and follows from the fact that
the IR divergent terms have the form of a universal factor
multiplying the one--photon exchange cross section, which does
not exhibit a spin dependence. Note that the cancellation
of IR divergences takes place only in the combination
Eq.~(\ref{f_combination}); the two--photon exchange contribution
to the individual functions $f_1$ and $f_2$
(even their imaginary parts) is IR divergent.

The invariant integral in Eq.~(\ref{Im_f1_f2_integral}) can be evaluated 
in an arbitrary reference frame. A convenient way is to convert it 
to a phase space integral over the intermediate electron momentum, 
which can be evaluated in the $ep$ center--of--mass (CM) frame using 
standard techniques. The relation of the CM momentum, $l_{\text{cm}}$, 
and scattering angle, $\theta_{\text{cm}}$, to the invariants 
$s$ and $t$ is (we assume zero electron mass)
\be
l_{\text{cm}} &=& \frac{s - M^2}{2\sqrt{s}} , \\
\sin^2 (\theta_{\text{cm}}/2) &=& \frac{-st}{(s - M^2)^2} .
\label{theta_from_t}
\ee
Evaluating the integral in this way we obtain a simple
result for the normal spin difference of the cross section,
\beq
\text{Im} \, (f_1^\ast f_2) \;\; = \;\; \frac{-e^6}
{256\, \pi \, l_{\text{cm}}^4 s^2 \sin^2\theta_{\text{cm}}} .
\label{Im_f1_f2_final}
\eeq
The normal spin asymmetry for the pointlike proton is then obtained
by multiplying with the kinematic factor of Eq.~(\ref{X_N_from_f}),
and dividing the result by the spin sum of the cross section, 
evaluated in one--photon exchange approximation, Eq.~(\ref{X_U_a}).
In terms of the CM variables,
\be
A_N &=& - \frac{2 \alpha \, l_{\text{cm}}^2 M}{s^{3/2}}
\nonumber \\
&\times&
\frac{\sin^3 (\theta_{\text{cm}}/2) \;  \cos (\theta_{\text{cm}}/2)} 
{\cos^2(\theta_{\text{cm}}/2) + (2 l_{\text{cm}}^2/s) \, 
\sin^4 (\theta_{\text{cm}}/2)} . 
\;\;
\label{A_N_final_cm}
\ee
Here $\alpha = e^2/(4\pi) = 1/137$ is the fine structure constant.
This result agrees with the one obtained earlier in 
Ref.~\cite{Barut60}; see also Ref.~\cite{Afanasev:2005mp}.
In particular, in the high--energy limit, $s \gg M^2$, one has 
$l_{\textrm{cm}} \approx \sqrt{s}/2$, and Eq.~(\ref{A_N_final_cm})
simplifies to
\beq
A_N \;\; = \;\; - \frac{\alpha M}{2 \sqrt{s}} \;
\frac{\sin^3 (\theta_{\text{cm}}/2) \;  \cos (\theta_{\text{cm}}/2)} 
{\cos^2(\theta_{\text{cm}}/2) + \frac{1}{2}
\sin^4 (\theta_{\text{cm}}/2)} ,
\label{A_N_simplified}
\eeq
which has its maximum at $\theta_{\textrm{cm}} = 2.18 = 125^\circ$. 

The sign of the normal spin asymmetry Eq.~(\ref{A_N_final_cm}) 
is what one expects from the simple picture of an electron scattering 
from a pointlike magnetic dipole, see Fig.~\ref{fig:xyz}b. In this 
picture the asymmetry is caused by the Lorentz force experienced by the 
charged particle moving in the magnetic field of the dipole, which in 
the scattering plane points in the direction opposite to the 
magnetic moment. As can be seen from Fig.~\ref{fig:xyz}b,
if the proton with magnetic moment $\bm{\mu}_p = e \bm{S} /(2M)$ 
is polarized upward, the electron with charge $-e$ is deflected to 
the right, leading to $A_N < 0$, \textit{cf.}\ Eq.~(\ref{A_left_right}).

We can use the result of the pointlike proton approximation to make
a rough order--of--magnitude estimate of the asymmetry expected 
in DIS experiments.
Figure~\ref{fig:an_tdep} shows the asymmetry for $s = 10$ and 
$20 \,\text{GeV}^2$, corresponding approximately to the values 
reached in $ep$ scattering at JLab with 6 and 12 GeV beam energy.
The asymmetry is shown both as a function of the CM scattering 
angle, $\theta_{\text{cm}}$, Eq.~(\ref{theta_from_t}),
and as a function of $|t| = -q^2 = Q^2$ itself. One sees that the asymmetry 
in this approximation is of the order of several times $10^{-4}$. 
The maximum value of the asymmetry, as well as its position
in $\theta_{\text{cm}}$, depend only weakly on $s$.
The change of the $t$--dependence with $s$ mostly reflects
the transformation from the kinematic variable $\theta_{\text{cm}}$
to $t$.

\begin{figure}
\includegraphics[width=7cm]{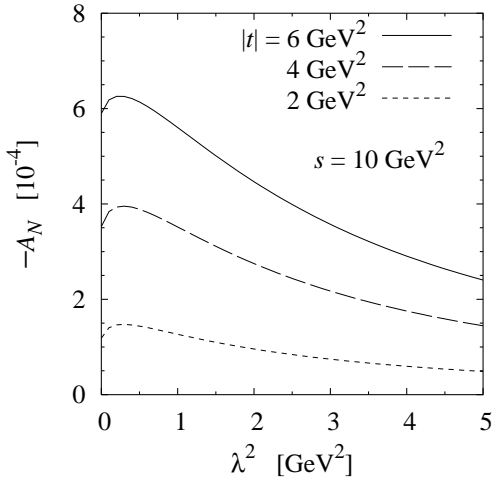}
\caption[]{The normal spin asymmetry $A_N$ (note the minus sign on the
axis) as a function of the photon mass, 
\textit{cf.}\ Eqs.~(\ref{Im_f1_f2_integral})
and (\ref{lambda_introduced}). The photon mass dependence
gives an indication of the average virtualities in the two--photon
box graph.}
\label{fig:lambda}
\end{figure}

It is interesting to study the distribution of photon 
virtualities in the integral (\ref{Im_f1_f2_integral}). This provides
information about the effective range of the two--photon exchange
interaction in the asymmetry, which will be important for 
the calculation of the asymmetry for a composite target
in Sec.~\ref{sec:composite}. It turns out that the distribution
of photon virtualities in the integral (\ref{Im_f1_f2_integral}) is
governed by the scales $Q^2 = -t$ and $s$ and does not involve any
extraneous scales, as a result of the IR finiteness of the integral. 
One way to illustrate this is by evaluating the integral 
(\ref{Im_f1_f2_integral}) with a non--zero ``photon mass,'' replacing the 
photon propagators by
\beq
\frac{1}{q_{1,2}^2} \;\; \rightarrow \;\;
\frac{1}{q_{1,2}^2 - \lambda^2} .
\label{lambda_introduced}
\eeq
The variation of the result with $\lambda^2$ gives an indication
of the effective photon virtualities in the loop.
Figure~\ref{fig:lambda} shows the asymmetry for $s = 10\, \text{GeV}^2$
as a function of the photon mass, $\lambda^2$, for several values of $t$,
corresponding to large--angle scattering ($\theta_{\text{cm}}$ not
close to $0$ or $\pi$). One sees that the photon mass dependence is very 
smooth, confirming that no large contributions arise from 
the region $|q_1^2|, |q_2^2| \ll Q^2$. We observe that for 
$\lambda^2 \gtrsim 1 \, \text{GeV}^2$ the $\lambda$ dependence is
well described by the form 
\beq
\text{Im} \, (f_1^\ast f_2) \;\; \propto \;\; 
\frac{1}{(Q^2_{\text{eff}} + \lambda^2)^2} ,
\eeq
which would be the dependence if the virtualities in the integral
were  ``frozen'' at $-q_1^2 = -q_2^2 = Q^2_{\text{eff}}$. Extracting 
the value of $Q^2_{\text{eff}}$ from a fit to the numerical results 
we find $Q^2_{\text{eff}} = (5.6, \, 5.8, \, 6.1) \, \text{GeV}^2$ 
for $Q^2 = (2, \, 4, \, 6) \, \text{GeV}^2$ in the given kinematics.
Note, however, that the effective virtuality thus estimated depends
strongly on the prescription; since the integrand of 
Eq.~(\ref{Im_f1_f2_integral}) is not positive definite, 
any definition of average is inherently ambiguous.
%
%

Another way of studying the distribution of photon virtualities in
the asymmetry is to represent the integral (\ref{Im_f1_f2_integral})
as an integral over one of the photon virtualities. This can be 
done using the fact that in the CM frame the virtuality $q_1^2$
is directly related to the angle between the initial and intermediate
electron momenta, $Q_1^2 \equiv -q_1^2 = 2 \, l_{\text{cm}}^2
[1 - \cos\theta (\bm{l}_1, \bm{l})]$. Integrating over the corresponding
azimuthal angle, one obtains a representation of the form
\beq
\text{Im} \, (f_1^\ast f_2) \;\; = \;\; \int_0^{2\, l_{\text{cm}}^2} 
dQ_1^2 \; F(Q_1^2) ,
\eeq
where the integrand turns out to be a piecewise constant function,
\beq
F(Q_1^2) \;\; = \;\; 
\left\{
\begin{array}{lc} C_1 \hspace{2em} & 0 < Q_1^2 < Q^2,
\\[2ex]
C_2 & Q^2 < Q_1^2 < 2\, l_{\text{cm}}^2 ,
\end{array}
\right.
\eeq
in which $C_1 < 0, C_2 > 0$, with values depending on $s$ and $Q^2$.
One sees that the characteristic scales in the distribution of virtualities
are $Q^2$ and $2 l_{\text{cm}}^2 \sim s$. Numerical studies show that
for large--angle scattering ($\theta_{\text{cm}}$ not
close to $0$ or $\pi$) the cancellation between the low and high 
virtuality regions is not precarious, and that the sign of the 
resulting integral is always determined by the high--virtuality 
contribution. This again proves that in the kinematics of large $s$ 
and $Q^2$ the contribution from virtualities $Q_1^2 \ll Q^2$ does 
not significantly change the result.

To complete our discussion of the transverse spin dependence 
of inclusive scattering from a point particle we need 
to comment also on the real photon emission (bremsstrahlung) channel,
$ep \rightarrow e'p\gamma$, the amplitudes for which are given by
the diagrams of Fig.~\ref{fig:point}d and e. The spin dependence
of the cross section in this channel can be discussed along the
lines of Eq.~(\ref{M2_from_X}) \textit{et seq.}, the only difference
being that the sum over final states includes the integration over 
the relative momenta of the three--body final state and the sum over
the photon polarizations. An expression analogous to 
Eq.~(\ref{X_N_from_helicity}) can be derived in terms of the
helicity amplitudes; however, since the diagrams (d) and (e) do not 
have an absorptive part (the intermediate particles are always 
off mass--shell) the helicity amplitudes are real and no transverse 
spin dependence is obtained. In this sense the pointlike target 
provides a model for fully inclusive scattering; only the bremsstrahlung
channel happens not to contribute in this case. Note that this is 
specific to scattering from a point particle; for a target with 
internal excitations the Compton amplitude has an absorptive part
and a non--zero interference cross section can in principle arise. 
This contribution is IR finite (\textit{cf.}\ the discussion
in Sec.~\ref{sec:divergences}) and thus can be discussed separately 
from two--photon exchange.
\section{Cancellation of infrared and collinear divergences}
\label{sec:divergences}
A new feature of two--photon exchange processes compared to one--photon
exchange is that divergent terms can appear in the scattering amplitude, 
related to the vanishing of the virtualities of (at least) one of the photons.
However, these divergent terms must cancel in the final result for
physical observables. The point particle calculation of Sec.~\ref{sec:point}
shows explicitly that IR divergences are absent in the spin--dependent 
part of the two--photon exchange cross section. We now want to demonstrate 
that this result is general and applies also to a target with internal 
structure; it follows from the general factorization property of IR 
singularities in QED. Furthermore, we analyze the collinear divergences 
which appear in the calculation of two--photon exchange corrections
in models of hadron structure with off mass--shell constituents,
and show that they cancel due to electromagnetic gauge invariance.
These results will be used in our studies of the transverse spin dependence
of the inclusive cross section in the presence of hadron structure below.

%
%
\begin{figure}
\includegraphics[width=3cm]{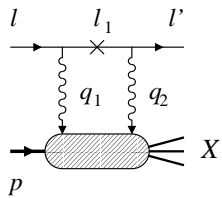}
\caption{The two--photon exchange amplitude giving rise to
a transverse spin dependence of the $eN \rightarrow e'X$ inclusive 
cross section.}
\label{fig:compton} 
\end{figure}
Consider the invariant amplitude for the two--photon transition 
to a (unspecified) hadronic final state, $eN\to e'X$, which enters 
in the transverse spin--dependent part of the inclusive $eN$ cross section, 
see Fig.~\ref{fig:compton} (\textit{cf.} Fig.~\ref{fig:twogamma}a). 
The absorptive part, in which the intermediate electron is on mass--shell, 
can be represented as 
\beq
\text{Im}\; \mathcal{M}_{ep\to e'X} \;\; = \;\; 
\int\frac{d^3l_1}{2E_1(2\pi)^3}\frac{e^4 \, l^{\mu\nu} \;
\text{Im}\, T_{\mu\nu}}{q_1^2 q_2^2} , 
\label{Im_M_integral}
\eeq
where $E_1$ is the energy of the intermediate electron, and
$q_{1,2}$ are the photon 4--momenta. In the numerator,
\begin{equation}
l_{\mu\nu} \;\; = \;\; 
\bar{u}(l') \gamma_\mu \hat{l}_1 \gamma_\nu u(l)
\label{l_mu_nu_def} 
\end{equation}
is the residue of the direct term of the electron virtual Compton 
amplitude (we neglect the electron mass), and $\text{Im} \, T_{\mu\nu}$
denotes the absorptive part of the virtual hadronic Compton amplitude for 
the $N \rightarrow X$ transition \cite{foot:absorptive}. As a consequence
of electromagnetic current conservation, the tensors satisfy the 
transversality conditions
\be
q_{2\mu} \, l^{\mu\nu} &=& 0, \hspace{2em}
l^{\mu\nu} \, q_{1\nu} \;\; = \;\; 0 ,
\label{transversality_leptonic}
\\ 
q_{2\mu} \, T^{\mu\nu} &=& 0, \hspace{2em} 
T^{\mu\nu} \, q_{1\nu} \;\; = \;\; 0 .
\label{transversality_hadronic}
\ee

The integral Eq.~(\ref{Im_M_integral}) can become divergent if either of the
exchanged photon virtualities, $q_1^2$ or $q_2^2$, tends to zero 
in parts of the integration region. One distinguishes two types 
of such singularities:
\[
\begin{array}{rcl}
q_1 \rightarrow 0, \;\; q_2 \rightarrow q  & \hspace{2em} 
&\textrm{``infrared,''}   \\[2ex]
q_1^2 \rightarrow 0$ \; with \; $q_1 \neq 0 & &\textrm{``collinear,''} 
\end{array}
\]
and likewise for $q_1 \leftrightarrow q_2$. The mechanism for the 
cancellation of these singularities in physical observables is 
quite different in the two cases.

The cancellation of IR singularities is governed by
the soft--photon theorem \cite{Low58}, which states that 
photons of wavelength $\lambda \gg R_{\text{hadron}}$ ``see''
only the charge and momenta of the initial and final particles in a
reaction, not their polarization or the details of the reaction mechanism. 
Using the method of Refs.~\cite{YFS61,Tsai61,Cahn:1970gp}, 
the IR divergent contributions
of individual two--photon exchange diagrams can be represented in
the form of a divergent factor, depending only on the charges and momenta 
of the initial and final particles, multiplying the one--photon exchange 
amplitude for the process. Because this factor is 
spin--independent, the IR divergent term in the spin--dependent 
cross section difference comes in the form of an overall factor 
multiplying the spin--dependent cross section difference in one--photon
exchange approximation, which is zero on grounds of the Christ--Lee
theorem. Note that the cross section for each individual target
polarization does have IR divergent terms; they cancel only at the
level of the cross section difference. This is exemplified by the
point particle calculation of Sec.~\ref{sec:point}, where one can 
verify that the two--photon exchange contribution to the absorptive parts 
of the individual amplitudes $f_1$ and $f_2$ are divergent, while the 
spin difference $\text{Im} \, (f_1^\ast f_2)$ is divergence--free. 
In summary, the reason why the cross section spin difference is IR finite 
is the spin--independence of soft--photon contributions.

The cancellation of collinear singularities in the two--photon exchange 
contribution to inclusive $eN$ scattering is due to the transversality
of the electron and hadron Compton tensors (related to 
electromagnetic gauge invariance), and happens already at
the level of the amplitude for given target spin. Physically, 
collinear singularities correspond to the emission of a finite--energy 
photon along the direction of the initial or final electron, which is
assumed to be strictly massless here ($m = 0$). Consider the case that 
the photon with $q_1$ is emitted along the direction of the initial electron
with 4--momentum $l$. The relevant integration region can be parametrized 
covariantly as
\be
l_1 &=& zl, 
\label{collinear_limit_1} 
\\
q_1 &=& l - l_1 \;\; = \;\; (1 - z) \, l,
\label{collinear_limit_2} 
\ee
where $z$ is the fraction of the momentum $l$ carried by the
intermediate electron. Obviously $q_1^2 = (1 - z)^2 l^2 = 0$
for massless electrons, and one encounters a divergence
if values $z \neq 1$ are kinematically allowed, as is generally true
in inelastic scattering (the case of elastic scattering will be 
discussed separately below). The only way a divergence can be avoided 
is if the numerator of the integral vanishes in the collinear limit,
Eqs.~(\ref{collinear_limit_1}, \ref{collinear_limit_2}).

Let us inspect the numerator of Eq.~(\ref{Im_M_integral}) in the collinear 
limit, Eqs.~(\ref{collinear_limit_1}, \ref{collinear_limit_2}). Using the 
anticommutation relations for the gamma matrices and the Dirac equation 
for the initial electron spinor, the tensor Eq.~(\ref{l_mu_nu_def}) can be 
brought into the form
\beq
\lim_{l_1 \rightarrow z l} l_{\mu\nu} \;\; = \;\;
\frac{2z}{1-z} \; j_\mu (l', l) \; q_{1\nu} ,
\label{l_mu_nu_collinear}
\eeq
where 
\beq
j_\mu (l', l) \;\; = \;\; \bar{u}(l') \gamma_\mu u(l)
\eeq
is the matrix element of the electromagnetic current between the
initial and final electron states. The contraction with the hadronic 
Compton tensor then gives zero by virtue of the transversality
condition Eq.~(\ref{transversality_hadronic}),
\beq
l^{\mu\nu} T_{\mu\nu} \;\; \propto \;\;
j^\mu (l', l) \, T_{\mu\nu} \, q_1^\nu \;\; = \;\; 0.
\eeq
A similar argument applies if the other photon momentum, $q_2$, 
becomes collinear to the final electron momentum, $l'$. 
In both collinear regions, the numerator in the integrand 
of Eq.~(\ref{Im_M_integral}) tends to zero simultaneously 
with the denominator and the integral becomes convergent. In summary,
the absence of collinear divergences in the two--photon contribution
to inelastic $eN$ scattering is directly related to the transversality of 
the hadronic Compton tensor. A similar observation was made earlier in 
Ref.~\cite{Afanasev:2004hp} in applications to the single-spin asymmetry 
of elastic $ep$ scattering induced by two--photon exchange.

The case of a pointlike target considered in Sec.~\ref{sec:point}
is somewhat special in the context of the above discussion.
For elastic scattering from a point particle, the only way 
in which $q_1^2$ could vanish is if the 4--vector $q_1$ tends to
zero, \textit{i.e.}, the only kinematically allowed value of $z$ 
in Eqs.~(\ref{collinear_limit_1}, \ref{collinear_limit_2}) is $z = 1$.  
In this case the collinear region is kinematically forbidden; 
the only singularities are IR divergences, which cancel by the
mechanism described earlier.

The issue of collinear singularities becomes critical when
one tries to incorporate effects of hadron structure in inelastic
$eN$ scattering with two--photon exchange. Specifically, in models
where the two--photon exchange couples to hadronic constituents which 
are off mass--shell, collinear divergences appear, which are canceled
only by contributions involving explicitly the interactions between
the constituents. This is because only the combination
of off--shell and interaction effects maintains electromagnetic gauge
invariance and transversality of the hadronic tensor. We note that
a recent calculation of the transverse spin asymmetry in inclusive DIS 
in the parton model \cite{Metz:2006pe}, which considered two--photon 
exchange with off--shell quarks without accompanying interaction effects, 
found a divergent result for the asymmetry \cite{foot:metz}. The arguments
presented above indicate that the reason for the divergence is the 
violation of electromagnetic gauge invariance in that approximation,
and point out what needs to be done to obtain a meaningful finite result.

%
%
\begin{figure}[t]
\includegraphics[width=8cm]{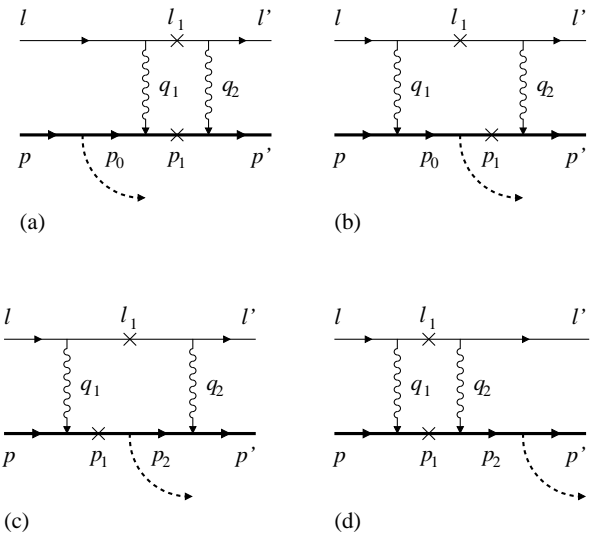}
\caption{The field--theoretical model for inelastic
electron scattering with two--photon exchange. Thin solid lines denote
the electron, thick solid lines the spinor particle (charge $+e$), 
dashed lines the massive neutral scalar particle. The crosses indicate
that the intermediate particle is on mass--shell (Cutkosky cut).}
\label{fig:ftmodel}
\end{figure}
To illustrate the point, we consider the simple field--theoretical model 
of an electron scattering from a pointlike spin--1/2 particle 
(charge $+e$), coupled to a massive neutral scalar particle
with a Lagrangian density
\beq
L_{\text{int}} \;\; = \;\; g \bar\Psi \Psi \phi .
\eeq 
This extension of the point particle calculation of Sec.~\ref{sec:point}
offers the simplest setting which allows one to study inelastic 
scattering with non--trivial target structure and includes both off--shell 
and interaction effects. 
We consider the scattering amplitude for the process in which 
a single scalar particle is produced in the final state. 
Its absorptive part is given by the sum of the cut Feynman diagrams 
of Fig.~\ref{fig:ftmodel} \cite{foot:ftmodel}. It can be shown explicitly
that the Compton tensor $T_{\mu\nu}$ defined by this model
satisfies the transversality conditions Eq.~(\ref{transversality_hadronic}),
provided that the full set of diagrams in Fig.~\ref{fig:ftmodel} is included. 
Obviously, the separate contributions from each of the diagrams
are not transverse.

Let us consider the diagram of Fig.~\ref{fig:ftmodel}a, where both photons 
couple to the spinor particle after emission of the scalar. 
When taken alone, the contribution of this diagram to the hadronic 
Compton tensor $T_{\mu\nu}$ is not transverse, which is related to 
the fact that the intermediate--state spinor particle with momentum 
$p_0$ is off mass--shell. The corresponding absorptive part reads
\be
\text{Im} \, T_{\mu\nu} &=& g \, \pi\delta(p_1^2 - M^2)
\nonumber \\
&\times& \bar{U}(p') \gamma_\mu 
(\hat{p}_1 + M) \gamma_\nu \frac{\hat{p}_0 + M}{p_0^2 -M^2} U(p) ,
\label{T_mu_nu_model_a}
\ee
where the delta function results from the Cutkosky cut. Since the spinor 
particle before the $q_1$ photon vertex is off mass--shell, 
$p_0^2 \neq M^2$, and the spinor particle after the vertex
is on mass--shell, $p_1^2 = M^2$, the $q_1$ photon may have zero
virtuality, $q_1^2 = 0$, while carrying a non--vanishing momentum. 
(For the $q_2$ photon both spinor particles at the vertex are
on mass--shell, and its virtuality can only go to zero if 
$q_2 \rightarrow 0$.) Consider now the collinear limit for the
$q_1$ photon, Eqs.~(\ref{collinear_limit_1}, \ref{collinear_limit_2}). 
In this case $q_1^2\rightarrow 0$, and from $q_2 = q - (1 - z)l$
(where $q = l - l'$) one obtains $q_2^2 = z q^2$. The fraction $z$ 
is kinematically fixed to be
\beq
z \;\; = \;\; (s_0 - M^2)/(s_0 - p_0^2) , 
\eeq
where 
\beq
s_0 \;\; \equiv \;\; (l + p_0)^2  \;\; = \;\; (l' + p')^2 
\eeq
is the invariant of the $q_1$ exchange subprocess, and the 
off--shell momentum $p_0$ is reconstructed from 4--momentum conservation, 
$p_0 = p' - q = p' - l + l'$. Likewise, the 4--momentum $p_1$ is
fixed as $p_1 = p' - q_2 = p' - z l + l'$. Calculating then the contraction 
of the electron and hadron Compton tensors in the collinear approximation,
Eqs.~(\ref{l_mu_nu_collinear}) and (\ref{T_mu_nu_model_a}), we find that it
does not vanish in the collinear limit, and that the integral
Eq.~(\ref{Im_M_integral}) is divergent. Using the fact that the delta
function reduces the momentum integral to a two--dimensional one, the 
divergent part can conveniently be calculated by converting 
Eq.~(\ref{Im_M_integral}) to an angular integral in the CM frame of the 
final electron and the spinor particle (\textit{cf}.\ Sec.~\ref{sec:point}).
Without the numerator, the divergent part is
\be
\lefteqn{\lim_{l_1 \rightarrow zl} \int \frac{d \Omega_1}
{(q_1^2-\lambda^2)(q_2^2-\lambda^2)} } &&
\nonumber \\
&=& -\frac{\pi}{q^2 E_1^2} \; \ln\left(\frac{4E_1^2}{z \lambda^2}\right) ,
\ee
where 
\beq
E_1 \;\; = \;\; \frac{s_0 - M^2}{2\sqrt{s_0}}
\eeq
is the energy of the intermediate electron in the CM frame,
and we have introduced a ``photon mass'' $\lambda$ to regularize
the singularity. Altogether, including the numerator factors,
we obtain for the contribution to the $ep \rightarrow e'X$ amplitude 
from the collinear region in diagram Fig.~\ref{fig:ftmodel}a
\be
\text{Im} \, \mathcal{M}_{ep\rightarrow e'X}
&=& \frac{e^4 g}{32 \, \pi \, q^2 \, E_1 \, \sqrt{s_0}} \; j_\mu (l', l)
\nonumber \\
&\times & \bar U(p')
\gamma_\mu (\hat{p}_1 + M) U(p) 
\nonumber \\
&\times & \ln\left(\frac{4E_1^2}{z \lambda^2}\right) ,
\hspace{1em}
\label{divergence_model_a}
\ee
where $p_1 = p' - z l + l'$. This expression is divergent in the physical 
limit, $\lambda \rightarrow 0$. Computing the interference cross section
with the one--photon exchange amplitudes for the same ``spinor + scalar''
final state, it is straightforward to verify that 
Eq.~(\ref{divergence_model_a}) leads to a divergent result for the 
transverse spin--dependent cross section in this model, similar to what 
was obtained in Ref.~\cite{Metz:2006pe}.

Following the arguments present above, the divergence resulting
from the ``off--shell'' diagram Fig.~\ref{fig:ftmodel}a should be 
canceled by the contribution of the ``interaction'' diagram
Fig.~\ref{fig:ftmodel}b, in which the spinor particle emits the
scalar between the two photon couplings. Indeed, we find that diagram 
Fig.~\ref{fig:ftmodel}b, when calculated in the same collinear--photon 
approximation, produces an expression equal to Eq.~(\ref{divergence_model_a}) 
but opposite in sign, leading to exact cancellation of the divergence in the 
resulting amplitude. Similarly, the divergent term arising from the second 
exchanged photon being collinear to the final--state electron ($q_2^2=0$) 
cancels in the sum of contributions from diagram Fig.~\ref{fig:ftmodel}d and
the diagram Fig.~\ref{fig:ftmodel}c. In summary, the field--theoretical
model explicitly demonstrates that collinear divergences are absent if 
off--shell and interaction effects are treated consistently and 
electromagnetic gauge invariance is maintained by the approximations. 
This observation serves as a basis of our studies of the spin--dependent 
two--photon exchange cross section in QCD and a constituent quark model 
in Secs.~\ref{sec:qcd} and \ref{sec:composite}.

The analysis presented here applies to collinear singularities 
arising from exchanges in which the virtuality of one of the 
photons tends to zero, while that of the other photon remains
non--zero. In general, collinear divergences can also arise from 
the region in which both photon virtualities tend to zero 
simultaneously; this case corresponds to vanishing momentum
of the intermediate electron in the CM frame. In our field--theoretical 
model such exchanges do not occur in the amplitudes for single--boson
emission into the final state (\textit{i.e.}, in first order of the 
coupling constant), because in all diagrams of Fig.~\ref{fig:ftmodel}
at least one of the internal spinor particles attached to a photon vertex 
is on mass--shell, making it impossible for that photon to have 
zero virtuality. They would, however, occur in higher--order 
amplitudes with multiple boson emission. Such exchanges would
give rise to $\ln^2$--type singularities in the individual diagrams,
which again cancel in the sum of all diagrams because of electromagnetic 
gauge invariance, as outlined above.

A comment is in order concerning the role of the electron mass
in collinear singularities. The above expressions were derived 
for the case of zero electron mass, $m = 0$. If the electron mass is not 
neglected, the electron polarization vector $\bm{s}$ can have a component 
transverse to the direction of the initial electron, and a new kind of 
transverse spin dependence of the inclusive $eN$ cross section appears, 
through a term proportional to
\beq
(\bm{s}, \bm{l} \times \bm{l}') .
\eeq
It corresponds to a beam spin asymmetry for electrons polarized in 
the direction normal to the scattering plane, while the target 
is unpolarized. For electrons polarized in this way, the electron
virtual Compton tensor in the collinear limit is no longer proportional 
to the collinear photon momentum as in Eq.~(\ref{l_mu_nu_collinear}), 
and collinear photon exchange makes a non--zero contribution to the 
beam spin--dependent cross section Ref.~\cite{Afanasev:2004hp}. 
In this case, however, the photon virtualities are limited by the 
(small) electron mass, and collinear photon exchange does not lead to 
a divergence but to a sizable finite contribution to the
beam spin--dependent cross section, which is enhanced by logarithmic
and double--logarithmic factors $\ln (Q^2/m^2)$ and $\ln^2 (Q^2/m^2)$, 
as was shown in Ref.~\cite{Afanasev:2004hp} 
(see also Refs.~\cite{Gorchtein:2005,Borisyuk:2005}).
\section{Transverse spin dependence in deep--inelastic scattering}
\label{sec:qcd}
The arguments of Sec.~\ref{sec:divergences} suggest that the
two--photon exchange in the transverse spin--dependent
cross section is free of QED IR and collinear divergences 
even when allowing for a non--trivial structure of the target and
the hadronic final state. Based on these findings, we now
want to discuss the transverse target spin dependence in DIS kinematics
in QCD. We are not aiming for a full calculation of the two--photon 
exchange contribution in the collinear factorization scheme. Rather, 
we want to discuss the underlying assumptions and ingredients in such 
a calculation, and prepare the ground for a self--consistent approximate 
treatment of the problem.

\begin{figure}
\includegraphics[width=5.5cm]{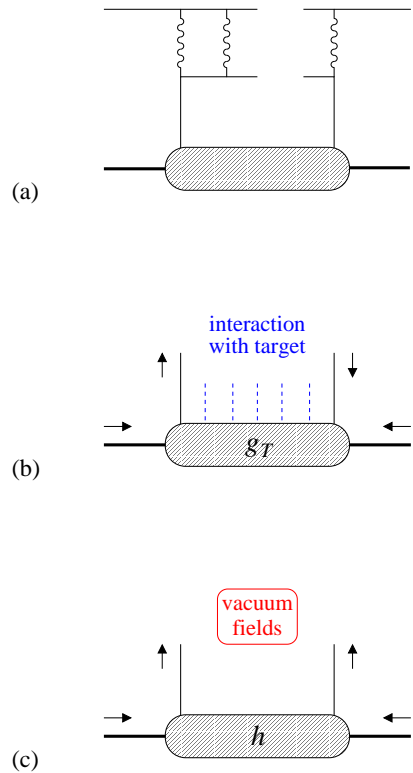}
\caption[]{Transverse spin dependence of the DIS cross section 
in QCD. (a) Assumption of dominance of two--photon exchange 
with the same quark. (b) Quark helicity--conserving process
involving interactions with the target remnants. 
(c) Quark helicity flip due to interaction with non--perturbative
vacuum fields.}
\label{fig:flip}
\end{figure}

Generally, in DIS kinematics we expect the dominant contribution 
to the target spin--dependent two--photon exchange cross section 
to arise from the amplitudes in which the two photons couple to 
a single quark, namely the same quark as is hit in the interfering 
one--photon exchange amplitude, see Fig.~\ref{fig:flip}a.
This follows from (a) the fact that the partonic final state in the
two--photon exchange amplitude needs to be the same as in the interfering 
one--photon exchange amplitude, (b) that no large contributions
arise from the soft regime of the two--photon exchange because of
the IR finiteness of the asymmetry. More precisely, the only way
in which a two--photon exchange coupling to different quarks could 
produce a final state similar to that of one--photon exchange in DIS 
would be if one of the photons were ``hard'' 
(with 4--momentum almost equal to $q$),
and the other were ``soft'' (with longitudinal and transverse 
momentum in the target rest frame of the order of the soft 
interaction scale, say, the inverse nucleon radius, $R_N^{-1}$). 
The amplitude of such ``hard--soft'' configurations in the two--photon 
exchange is not enhanced compared to average configurations, 
thanks to the overall IR finiteness of the process.
On the other hand, the phase space (integration volume) for
such configurations is suppressed compared to those in which
the two--photon exchange couples to the same quark and both photons
have ``average'' 4--momenta of the order $q/2$. Thus, the 
two--photon coupling to the same quark should dominate. 
(A more explicit version of this argument will be presented 
in Section~\ref{sec:soft} for the case of soft high--energy 
scattering, using closure over non--relativistic quark model states.)
While this conclusion seems plausible, we presently have no
way of proving it more rigorously, such as by way of a formal twist
expansion as in one--photon DIS. We shall adopt it as a working
assumption in the following.

%
%
It then follows that the transverse spin--dependent cross section
can be described in a ``parton--like'' picture, in which the reaction
happens predominantly with a single quark in the target. In this case 
one can easily see that one is dealing with two distinct contributions,
defined by whether the quark helicity is conserved or flipped in
the quark subprocess, see Fig.~\ref{fig:flip}b and c. [The hadron 
helicity is always flipped between the in and out state, as required 
by the transverse spin asymmetry, \textit{cf.} Eq.~(\ref{X_N_from_helicity}).]
Perturbative QCD interactions (gluon radiation) preserve the 
quark helicity and thus do not ``mix'' these contributions.

In the process of Fig.~\ref{fig:flip}b the quark helicity is conserved 
in the quark subprocess. This contribution to the transverse spin--dependent
cross section of unpolarized electron scattering is similar to that 
giving rise to the transverse spin structure function 
$g_T \equiv g_1 + g_2$ in longitudinally polarized electron scattering 
with a transversely polarized target. The latter is determined by
the matrix element of the quark helicity--conserving (chirally even), 
transversely polarized twist--3 density $g_T(x)$, 
defined as ($i = 1, 2$ is a transverse index)
\be
S^i \; g_{T, f}(x) &=& \frac{p^+}{M}
\int\frac{dz^-}{8\pi} \; e^{i x p^+ z^-/2} 
\nonumber \\
&\times& \langle p S_T \, | \; \bar\psi_f (0) 
\gamma^i \gamma_5 \psi_f (z) \; | \, p S_T 
\rangle_{\bm{z}_\perp = 0, \; z^+ = 0} ,
\nonumber \\
\label{g_T_quark}
\ee
where $z^{\pm} \equiv z^0 \pm z^3$ and $\bm{z}_\perp$ are the
usual light--cone vector components, $\bar\psi, \psi$ are the quark 
fields, $f$ denotes the quark flavor, and we have omitted the gauge 
link in the light--ray 
operator for brevity. Indeed, the QCD calculation of the quark 
helicity--conserving two--photon exchange contribution would
start from the ``handbag graph'' of Fig.~\ref{fig:flip}a, with
the quark density given by Eq.~(\ref{g_T_quark}) \cite{Metz:2006pe}.
However, keeping only this graph is not a consistent approximation.
On one hand, evaluating it with the initial and final quark on mass shell 
would give zero, as can be seen from the results of Sec.~\ref{sec:point},
which show explicitly that the spin--dependent interference term for 
an on--shell point particle is proportional to the particle mass.
On the other hand, allowing for finite virtuality of the initial and 
final quark leads to the appearance of collinear 
divergences \cite{foot:metz}, which are canceled only
by graphs with explicit interactions of the active quark in the 
intermediate and final state, as was shown in detail in 
Sect.~\ref{sec:divergences}. The quark helicity--conserving contribution 
of Fig.~\ref{fig:flip}b is thus of essentially ``non--partonic'' character, 
requiring interaction of the active quark with the spectator system.

In the process of Fig.~\ref{fig:flip}c the quark helicity is flipped
in the course of the electron--quark scattering process. In short--distance
processes such as DIS, the amplitude for quark helicity flip is
usually thought to be of the order of the current quark mass, 
$m_f \sim \textrm{few MeV}$, which is very small compared to 
typical hadronic mass scales. However, it is known that
at larger distances the phenomenon of spontaneous chiral symmetry 
breaking sets in, and helicity--flip amplitudes of the order of a
typical ``constituent quark'' mass, $M_q \approx 300 \, \textrm{MeV}$
are generated dynamically by interactions with non--perturbative
vacuum fluctuations. We suggest here that this phenomenon plays 
an important role in the transverse spin--dependent $eN$ cross section 
even in DIS kinematics. This perhaps somewhat surprising assertion 
is supported by the following arguments.

First, in QCD significant helicity--flip amplitudes should be present 
for quarks with virtualities smaller than some characteristic 
scale, $\mu^2_{\textrm{chiral}}$, which is determined by the 
typical size of the non--perturbative field configurations 
instrumental in the spontaneous breaking of chiral symmetry.
Quarks with virtualities $\gg \mu^2_{\textrm{chiral}}$ should 
experience only helicity--conserving perturbative interactions. 
This is explicitly seen in dynamical models of chiral symmetry 
breaking in QCD based on constituent quarks, such as the instanton 
vacuum or Dyson--Schwinger equations, which show a 
momentum--dependent dynamical quark mass which reduces to the
current quark mass at virtualities $\gg \mu^2_{\textrm{chiral}}$
(in the instanton vacuum the chiral symmetry breaking scale is 
determined by the average instanton size in the vacuum,
$\mu^2_{\textrm{chiral}} \sim \rho^{-2} \approx (0.6\, \textrm{GeV})^2$. 
Neglecting for the moment perturbative QCD radiation, the transverse 
spin dependence would be given by the imaginary part of the two--photon 
``box graph,'' in which the intermediate quark is on mass--shell.
It is precisely such low--virtuality quarks which experience large 
helicity--flip amplitudes due to chiral symmetry 
breaking \cite{foot:mass}.

Second, the previous argument can be generalized to account for the 
presence of perturbative QCD radiation. Generally, QCD radiation 
in DIS processes leads to a broad distribution of quark virtualities,
extending up to $Q^2$. The condition to propagate through a 
low--virtuality quark line (in order to enable a helicity flip) 
results in a suppression at the photon--quark vertices, measured
by the Sudakov form factor. In the usual DIS cross section,
which is given by the imaginary part of the quark Compton amplitude,
this suppression is compensated by real gluon emissions.
In the two--photon exchange process responsible for the transverse
spin asymmetry, it is likely that this compensation happens only 
incompletely, and that a residual Sudakov suppression remains.
To estimate the magnitude of this suppression, we consider the standard 
on--shell Sudakov form factor,
\beq
S(Q^2) \;\; = \;\; \exp\left( -\frac{\alpha_s C_F}{4\pi} \, \ln^2 
\frac{Q^2}{\mu^2} \right) ,
\eeq
where 
$\alpha_s = 4\pi / \left[ b \ln (Q^2 / \Lambda^2_{\textrm{QCD}}) \right]$ 
is the one--loop running coupling constant at the scale $Q^2$,
with $b = 11 - (2/3) N_f$ and $\Lambda^2_{\textrm{QCD}} = 0.20 \, 
\textrm{GeV}$ for $N_f = 3$, and $C_F = 4/3$. Furthermore, $\mu^2$ 
denotes the IR cutoff for gluon emission. In the light of the above
arguments about dynamical chiral symmetry breaking it is natural
to identify this cutoff with the chiral symmetry breaking scale,
\beq
\mu^2 \;\; \sim \;\; \mu^2_{\textrm{chiral}} .
\eeq
Gluons of virtualities $k^2 < \mu^2$ are regarded as part of the 
non--perturbative vacuum fluctuations which lead to the spontaneous 
breaking of chiral symmetry and thus ``contained'' in the dynamical
quark mass. Specifically, with the instanton vacuum value 
$\mu^2 = \rho^{-2} = 0.36\, \textrm{GeV}^2$ 
we obtain
\beq
S(Q^2) \;\; = \;\; (0.89, 0.86, 0.83)
\;\; \textrm{for} \;\; Q^2 \; = \; (2, 3, 4) \, \textrm{GeV}^2 .
\eeq
With this value of IR cutoff the Sudakov suppression of
low--virtuality quark lines is not very substantial \cite{foot:sudakov}. 
We conclude that a potentially sizable contribution to the transverse 
target spin dependence in inclusive DIS should come from the quark 
helicity--flip process of Fig.~\ref{fig:flip}c. If we chose instead 
the IR cutoff to be of the order
\beq
\mu^2 \;\; \sim \;\; \Lambda^2_{\textrm{QCD}} ,
\eeq
we would obtain a substantially larger Sudakov suppression. 
With $\mu^2 = \Lambda^2_{\textrm{QCD}}(N_f = 3) 
= 0.04 \, \textrm{GeV}^2$ we would find
\beq
S(Q^2) \;\; = \;\; (0.56, 0.52, 0.5)
\;\; \textrm{for} \;\; Q^2 \; = \; (2, 3, 4) \, \textrm{GeV}^2 .
\eeq
In this case amplitudes with quark helicity--flip would be
significantly suppressed in QCD compared to the constituent quark
model estimate. In the context of the present phenomenological 
discussion the choice of IR cutoff should in principle 
be regarded as an additional assumption; while it seems natural 
to choose it of the order of the chiral symmetry breaking scale, 
this could be rigorously justified only in an approximation scheme 
which treats the non--perturbative helicity-flipping fluctuations 
and perturbative gluon radiation in a unified framework.

In summary, we argue that a potentially sizable contribution to the
transverse target spin dependence in inclusive DIS results from
the quark helicity--flip process of the type Fig.~\ref{fig:flip}c.
This contribution is of the order of a typical ``constituent quark''
mass, $M_q \approx 300 \, \textrm{MeV}$, multiplying the twist--2
quark transversity distribution, which is defined as
($S$ denotes the nucleon polarization 4--vector)
\be
h_f (x) &=&  \int \frac{dz^-}{8\pi} \; e^{i\xi p^+ z^-/2} 
\nonumber \\
&\times & 
\langle p S_T | \; \bar\psi_f (0) \; \gamma^+\gamma_5 \hat S \;
\psi_f (z) \; | p S_T \rangle_{\bm{z}_\perp = 0, \; z^+ = 0} .
\nonumber \\
\label{h}
\ee
For a review of the properties of this distribution and 
its relation to other DIS observables, see \textit{e.g.}\ 
Ref.~\cite{Barone:2001sp}.

\begin{figure}
\includegraphics[width=7cm]{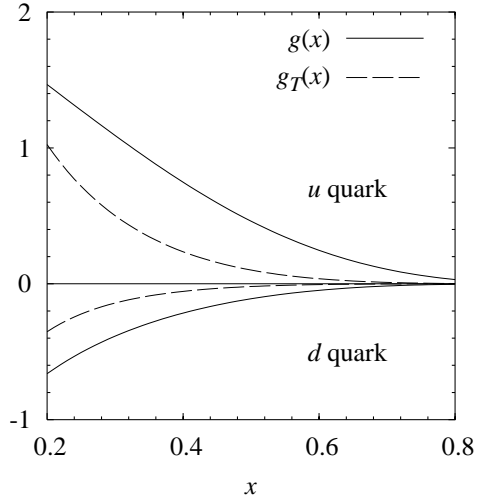}
\caption[]{The twist--3 transversely polarized quark distribution, $g_T(x)$, 
as estimated from Eq.~(\ref{g_T_from_ww}), evaluated with the
polarized parton densities of Ref.~\cite{Gluck:2000dy}.}
\label{fig:ww}
\end{figure}

It is interesting to compare the order--of--magnitude of the expected 
helicity--conserving and helicity--flip contributions to the spin--dependent 
cross section. While we can estimate the helicity--flip contribution in 
terms of the quark transversity distribution in the nucleon and the 
spin--dependent cross section for a pointlike constituent quark
(see Sec.~\ref{sec:composite}), we cannot presently calculate the
helicity--conserving contribution in terms of $g_{T, f}$ and twist--3 
quark--gluon operators in the nucleon. However, we can compare
the ingredients, $g_{T, f}$ and $h_f(x)$, and try to guess the
relative magnitude of the subprocess amplitudes in both contributions.
Using the Wandzura--Wilczek relation for $g_2$ \cite{Wandzura:1977qf}, 
which is valid in QCD up to terms proportional to twist--3 quark--gluon
operators, we can express $g_{T, f}$, Eq.~(\ref{g_T_quark}), as
\beq
g_{T, f} (x) \;\; = \;\; \int_x^1 \frac{dy}{y} g_f (y) 
\;\; + \;\; \textrm{quark--gluon},
\label{g_T_from_ww}
\eeq
where $g_f$ denotes the longitudinally polarized twist--2 
quark density. The part given by matrix elements of twist--3 
quark--gluon operators was measured in the SLAC E155 
\cite{Anthony:2002hy} and JLab Hall A \cite{Zheng:2004ce}
experiments and found to be small ($< 10^{-2}$), confirming 
theoretical predictions from the instanton vacuum model \cite{Balla:1997hf}. 
Neglecting it, we can calculate $g_{T, f}$ in terms of the twist--2 polarized 
parton densities. Figure~\ref{fig:ww} shows $g_T(x)$ as estimated 
from  Eq.~(\ref{g_T_from_ww}), using the polarized parton 
densities of Ref.~\cite{Gluck:2000dy}.
One sees that for $x \gtrsim 0.3$ the $g_{T, f}(x)$ are smaller 
than $g_f(x)$ at least by a factor of 2. 
%
%
A straightforward comparison between the helicity--flip and
helicity--conserving contributions, assuming that the amplitudes 
of the quark subprocesses are otherwise comparable, is then
\beq
M_q \, h_f(x) \;\; \leftrightarrow \;\; M \, g_{T, f}(x) ,
\label{h_g_T_comparison}
\eeq
where $M_q$ is a ``constituent quark'' mass, measuring the
generic strength of the quark helicity--flip amplitude due to 
non--perturbative vacuum fields. Since it is reasonable 
to assume that $h_f(x) \approx g_f(x)$, and $M_q$ is not small
(in the constituent quark model, $M_q \sim M/3$), we would conclude 
that the helicity--conserving contribution should be of the same 
order--of--magnitude as the helicity--flipping contributions.
At least one can say that substantially different values for the two 
contributions could result only if the electron--quark subprocess 
amplitudes are very different in the two cases.

In fact, one can argue that the comparison of the two contributions 
as in Eq.~(\ref{h_g_T_comparison}) overestimates the helicity--conserving 
contribution. Namely, the electron--quark scattering amplitude for the 
helicity--conserving process is zero for on--shell, collinear quarks,
and requires non--zero virtuality. A more realistic comparison would
thus be
\beq
M_q \, h_f(x) \;\; \leftrightarrow \;\; 
\frac{\langle \bm{k}_T^2 \rangle}{M} \; g_{T, f}(x) ,
\eeq
where $\langle \bm{k}_T^2 \rangle$ denotes the average transverse momentum
squared in the transverse momentum--dependent twist--3 distribution.
The factor $\langle \bm{k}_T^2 \rangle / M^2$ further reduces the
helicity--conserving compared to the helicity--flip contribution.

To summarize, we expect the dominant contribution to the transverse
spin--dependent DIS cross section to come from amplitudes in which
the two--photon exchange couples to a single quark. There are two distinct 
contributions to the transverse spin--dependent interference cross section 
in DIS, in which the quark helicity is either conserved or flipped in
the electron--quark subprocess. Both contributions are ``higher twist'' 
in the sense that they involve dynamical effects not present in the 
leading--twist approximation (explicit spectator interactions or 
quark masses). We have argued that the quark helicity--flip contribution 
in QCD should be sizable if Sudakov suppression
starts only at the chiral symmetry breaking scale,
$\mu^2_{\textrm{chiral}} \gg \Lambda^2_{\textrm{QCD}}$. 
Our order--of--magnitude estimates show that the quark--helicity 
conserving contribution is unlikely to dominate over the 
helicity--flip one. 
\section{Constituent quark model of composite nucleon}
\label{sec:composite}
We now want to make a quantitative estimate of the transverse spin
asymmetry of the DIS cross section which reflects the qualitative
findings of our analysis of Sec.~\ref{sec:qcd}. 
To this end, we employ a relativistic constituent quark model
in the light--front formulation. This framework offers proper
relativistic kinematics, while nevertheless maintaining close 
correspondence to the non--relativistic description of bound states
in the rest frame. The model describes DIS as elastic scattering from 
pointlike constituent quarks; it has a partonic limit, and the parton 
densities can be expressed as the longitudinal momentum densities 
of the light--cone wave functions (from the QCD point of view 
these correspond to the parton densities at a low normalization 
point, $\mu^2 \sim R_N^{-2}$). Most importantly, through the
constituent quark mass this model also generates non--zero 
quark helicity--flip amplitudes, which play an important role
in the transverse spin asymmetry (see Sec.~\ref{sec:qcd});
this aspect of constituent quark models was explored previously
in relation to the high--$Q^2$ behavior of the proton form factor 
ratio $Q F_2 / F_1$ \cite{Miller:2002qb}.

To arrive at a fully self--consistent scheme of approximations we endow
the constituent quark model with the additional dynamical assumption
that the nucleon is a weakly bound state (``composite''). That is, 
we suppose that the quark transverse momenta, which are of the
order of the inverse transverse size of the bound state, 
are parametrically small compared to the constituent quark mass,
\beq
\langle \bm{k}_T^2 \rangle \;\; \sim \;\; R_N^{-2} \;\; \ll \;\; M_q^2 . 
\label{neglect_k_T}
\eeq
This assumption permits several simplifications in the calculation
of the transverse spin--dependent cross section from two--photon exchange. 
First, it suppresses two--photon exchange with different quarks
and other interference contributions involving different quarks
beyond the generic suppression discussed in Sec.~\ref{sec:qcd} 
(see also Sec.~\ref{sec:soft}), and leaves the ``parton--like''
processes as the dominant ones. Second, among the ``parton--like''
processes it suppresses the quark helicity--conserving contribution 
of Fig.~\ref{fig:flip}b (which is proportional to 
$\langle \bm{k}_T^2 \rangle$) and leaves
the quark helicity--flip contribution of Fig.~\ref{fig:flip}c
(which is proportional to $M_q$) as the dominant one. Third, it allows 
us to consistently evaluate the latter in a relativistic impulse 
approximation with on--shell quarks, which exactly preserves 
electromagnetic gauge invariance and is free of collinear divergences.

The compositeness assumption (\ref{neglect_k_T}) is not intended as a 
reflection of actual nucleon structure (in reality $|\bm{k}_T| 
\sim \text{few 100 MeV}$ in the constituent quark model), but as a 
theoretical idealization which allows us to calculate the two--photon 
exchange cross section in a self--consistent scheme. Compared to the usual
one--photon exchange approximation for form factors and structure 
functions, in two--photon exchange processes one is dealing with several 
new effects (collinear divergences, exchanges with different constituents) 
which can qualitatively distort the results if not treated consistently. 
One therefore has to be prepared to make stronger assumptions about 
the structure of the bound state.

%
%
\begin{figure}
\includegraphics[width=5cm]{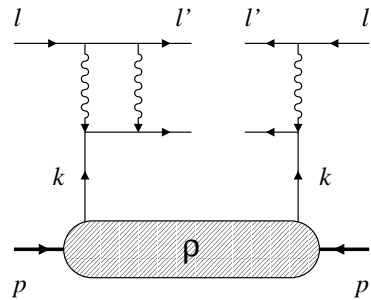}
\caption[]{Transverse spin--dependent two--photon exchange
cross section in the constituent quark model with the composite
nucleon approximation, Eq.~(\ref{impulse}).}
\label{fig:cqm}
\end{figure}
The technical implementation of the above ideas takes the form
of a relativistic impulse approximation, in which the electron
scatters elastically from a massive, on--shell constituent quark
quarks, see Fig.~\ref{fig:cqm} \cite{Miller:2004rc}. In this 
approximation the squared modulus of the invariant amplitude,
summed over all hadronic final states (corresponding to 
the cross section for inclusive $eN$ scattering) is given by
\be
\sum_X |\mathcal{M}_{eN \rightarrow e'X}|^2
&=& \sum_f
\int_0^1 \frac{d\xi}{\xi} \int \frac{d^2 \bm{k}_T}{(2\pi)^2}
\nonumber \\
&\times&
\text{tr} \left[ \, \rho_f (\xi, \bm{k}_T | p) 
\; \Gamma_f (k, l, l') \, \right] ,
\label{impulse}
\ee
where $k$ is the 4--momentum of the active quark, 
with light--cone 3--momentum components 
$k^+ \equiv k^0 + k^3 = \xi p^+$
and $\bm{k}_T$, and energy $k^- = (\bm{k}_T^2 + M_q^2)/ k^+$, 
corresponding to $k^2 = M_q^2$ (for simplicity we assume the
constituent quark mass $M_q$ to be the same for both light
quark flavors $f = u, d$). The matrix $\Gamma_f$ represents
the squared modulus of the invariant amplitude for elastic scattering 
of the electron from the on--shell quark with 4--momentum $k$,
\beq
|\mathcal{M}_{eq \rightarrow e'q}|^2
\;\; = \;\; 
\bar U_f (k) \; \Gamma_f (k, l, l') \; U_f(k) ,
\label{Gamma}
\eeq 
where $U_f$ denotes the initial quark spinor, and the dependence
on the initial quark helicity is contained in the spinors [the final
quark helicity is summed over, and the initial/final electron helicities 
are averaged/summed over, as in Eq.~(\ref{M2_from_X})]. Furthermore,
in Eq.~(\ref{impulse}) $\rho_f$ denotes the density matrix of the active 
quark in the nucleon, depending on the quark momentum variables 
$\xi$ and $\bm{k}_T$, as well as on the nucleon polarization. 
On grounds of Lorentz invariance, parity invariance, and the constraints 
imposed by the Dirac equation for the quark spinors, we can parametrize 
the density matrix for a quark in a transversely polarized nucleon as
\beq
\rho_f \;\; = \;\; \frac{\hat{k} + M_q}{2}
\left[ f_f (\xi, \bm{k}_T) - \gamma_5 \hat{a}
\; h_f (\xi, \bm{k}_T) \right] ,
\eeq
where
\beq
a \;\; \equiv \;\; S - \frac{(Sk) k}{k^2},
\hspace{2em} (ak) \;\; = \;\; 0
\eeq
is the component of the nucleon polarization vector orthogonal
to the quark 4--momentum. The functions $f_f (\xi, \bm{k}_T)$ and 
$h_f (\xi, \bm{k}_T)$ in this parametrization are related to the 
quark unpolarized and transversity parton densities in this 
model by
\be
\int \frac{d^2 k_T}{(2\pi)^2} \; f_f (\xi, \bm{k}_T) 
&=& f_f (\xi), \\
\int \frac{d^2 k_T}{(2\pi)^2} \; h_f (\xi, \bm{k}_T) 
&=& h_f (\xi).
\ee
By explicit calculation one can verify that the transversity
parton density thus obtained coincides with the one defined
in terms matrix elements of the quark light--ray operators,
Eq.~(\ref{h}), if the quark fields there are identified with the 
massive constituent quarks of this model. As already mentioned, 
in the context of QCD these parton densities would
correspond to a low normalization point of $\mu^2 \sim R_N^{-2}$.

When calculating DIS observables in the model defined by 
Eq.~(\ref{impulse}) \textit{et seq.}, we work in a frame where the 
initial proton momentum and the momentum transfer are collinear 
and have components along the $3$--direction, $\bm{q}_T, \bm{p}_T = 0$, 
with $p^+ > 0$. The mass--shell condition for the final--state quark,
$(k + q)^2 = M_q^2$, then fixes the plus momentum fraction
of the initial quark to be
\beq
\xi \;\; = \;\; x \; \frac{1 + \sqrt{1 + 4 M_q^2 / Q^2}}
{1 + \sqrt{1 + 4 x^2 M_{\phantom{q}}^2 / Q^2}},
\label{xi}
\eeq
where
\beq
x \;\; \equiv \;\; \frac{Q^2}{2 (p q)}
\eeq
is the usual Bjorken variable of the DIS process. Following the
composite nucleon assumption, we neglect corrections of the 
order $\bm{k}_T^2/s$ in the quark momentum fraction but retain 
corrections due to the finite proton and quark masses.
In this approximation the 4--momentum of the active quark can 
be expressed covariantly as 
\beq
k \;\; = \;\; A \, p \; + \; B \, q \; + \; k_\perp ,
\label{k_covariant}
\eeq
where $(p k_\perp), (q k_\perp) = 0$, and the scalars $A$ and $B$ 
are given by
\be
A &=& \frac{\xi (\eta Q^2 + 2 M_q^2)}{\eta Q^2 + M_q^2 + \xi^2 M^2},  
\label{A_sol} \\
B &=& 
\frac{\eta (M_q^2 - \xi^2 M^2)}{\eta Q^2 + M_q^2 + \xi^2 M^2} ,
\label{B_sol}  
\ee
with
\beq
\eta \;\; \equiv \;\; \frac{1}{2}
\left( 1 + \sqrt{1 + \frac{4 M_q^2}{Q^2}} \right) .
\eeq
The invariant energy of the electron--quark subprocess 
is then obtained as
\beq
s_{\text{sub}} \;\; \equiv \;\; 2(lk) \;\; 
= \;\; A (s - M^2) - B Q^2 + M_q^2 ,
\label{s_sub}
\eeq
up to terms proportional to the quark transverse momentum
which give corrections of the order $\langle \bm{k}_T^2\rangle$.

The scheme of approximation defined by Eqs.(\ref{xi})--(\ref{s_sub}) has 
several interesting properties. First, in the limit $Q^2 \rightarrow \infty$ 
we recover $\xi = x$ and $s_{\text{sub}} = xs$, as in the parton model.
Second, in the limit of zero binding, if we consider the nucleon as an
assembly of free quarks of mass $M_q$ and neglect the binding forces
between them, each quark should carry a fraction $M_q/M$ of the nucleon's 
momentum. Indeed, for $x = M_q/M$ Eq.~(\ref{xi}) gives $\xi = x = M_q/M$, 
and from Eqs.~(\ref{A_sol}) and (\ref{B_sol}) one obtains $A = x$ and 
$B = 0$, showing that our approximations respect this limit.
Third, our approximations are consistent with the overall kinematic 
boundaries of inclusive $eN$ scattering. For a given $eN$ CM energy
($s = 2 E_{\textrm{beam}} M + M^2$ for fixed--target experiments)
the minimum value of $x$ attainable is 
\beq
x_{\textrm{min}} \;\; = \;\; \frac{Q^2}{s - M^2 - Q^2 M^2 / (s - M^2)} ,
\label{x_min}
\eeq
corresponding to the maximum allowed energy loss of the electron, 
or a laboratory scattering angle of $\theta_{\textrm{lab}} = \pi$.
Conversely, for given $x$ the maximum attainable value of $Q^2$ is
\beq
Q^2_{\textrm{max}} \;\; = \;\; \frac{x (s - M^2)}{1 + x M^2 / (s - M^2)} .
\label{Q2_max}
\eeq
With our choice of kinematic variables for the electron--quark 
subprocess, Eqs.~(\ref{xi})--(\ref{s_sub}), this overall 
kinematic boundary corresponds exactly to the maximum value of 
the CM scattering angle of the electron--quark subprocess,
$\theta_{\textrm{cm}} \text{(electron-quark)} = \pi$, as one can show
by explicit calculation. The reason for this coincidence is that
the overall kinematic boundary corresponds to perfectly collinear 
kinematics (in the laboratory frame the electron bounces back with
zero transverse momentum), which is correctly described in our 
approximation where transverse momenta are neglected.

The calculation of the normal spin asymmetry of the $ep$ cross section
in the composite nucleon approximation defined above is straightforward, 
and essentially amounts to evaluating the asymmetry for the pointlike 
target in the kinematics of the quark subprocess. The matrix $\Gamma_f$
representing the squared amplitude of the quark--level 
subprocess, Eq.~(\ref{Gamma}), is given by
\beq
\Gamma_f \;\; = \;\; \frac{1}{2M_q} \left( 
X_{U, f} + \frac{\hat N_{\text{sub}} \gamma_5}{\sqrt{-N_{\text{sub}}^2}} 
\, X_{N, f}
\right) ,
\eeq
where $N_{\text{sub}}^\mu \equiv -4 \epsilon^{\mu\alpha\beta\gamma} 
l_\alpha l'_\beta k_\gamma$
is the normal 4--vector of the electron--quark subprocess
[\textit{cf}.\ Eqs.~(\ref{N}) and (\ref{k_covariant})], 
and the functions $X_{U, f}$ and
$X_{N, f}$ correspond to the results of Sec.~\ref{sec:point} with
$s \rightarrow s_{\text{sub}}$ [\textit{cf}.\ Eq.~(\ref{s_sub})], 
$M \rightarrow M_q$, and target charge $e \rightarrow e_f e$, 
where $e_f$ are the fractional quark charges. In our scheme of
approximation,where the quark transverse momenta are neglected
in kinematic factors, one has $(S N_{\text{sub}}) / \surd (-N_{\text{sub}}^2) 
\approx (SN) / \surd (-N^2)$. The result for the transverse spin 
asymmetry in inclusive DIS in the composite nucleon approximation
can then be expressed as
\beq
A_N (s, Q^2, x)_{\text{comp}} \;\; = \;\; 
R(\xi ) \; A_N (s_{\text{sub}}, Q^2)_{M = M_q} ,
\label{A_N_comp}
\eeq
where $\xi$ and $s_{\text{sub}}$ are given by Eqs.~(\ref{xi})
and (\ref{s_sub}), and $A_N$ on the right--hand side 
is the asymmetry for a pointlike constituent quark of charge 
$+e$ and mass $M_q$ [\textit{i.e.}, Eq.~(\ref{A_N_final_cm}) 
with the mass $M$ replaced by $M_q$], evaluated at the 
subprocess invariants $s_{\text{sub}}$ and $Q^2$.
The information about the quark structure of the target 
is contained in the structure factor 
\beq
R(\xi) \;\; \equiv \;\;
\frac{\sum_f e_f^3 \; h_f(\xi)}{\sum_f e_f^2 \; f_f (\xi)} ,
\label{R_def}
\eeq
which is the ratio of the sums of quark transversity 
and unpolarized parton densities, weighted with the quark charges 
corresponding to the two--photon -- one--photon interference cross 
section (numerator) and the one--photon cross section (denominator).
This ratio depends on the spin/flavor wave function of the
quark bound state, as well as on the momentum distribution
of the quarks. We shall discuss specific models for this
ratio in Sec.~\ref{sec:numerical}. 

Equation~(\ref{A_N_comp}) was derived in the approximation of weak 
binding between the constituents, where the quark momentum distributions
are concentrated around $\xi \sim M_q/M \sim 1/3$. It therefore should be 
applied only in the region around $x \sim 0.3$. In particular, for 
$x \rightarrow 1$ correlations between constituents in the wave function 
become important, and the picture of the composite nucleon is 
no longer applicable.

A cautionary remark is in order concerning the model dependence 
of the results presented here. The transverse spin--dependent DIS 
cross section involves not only the ``good'' ($+$) light--cone 
component of the current operator. This is seen \textit{e.g.}\ in the 
study of the high--energy behavior of the asymmetry 
(see Sec.~\ref{sec:soft}), where it is noted that the leading high--energy 
contribution to the cross section, resulting from $+$ the plus current 
components only, has no transverse spin dependence. Generally, in 
light--front quantization observables involving other than the ``good'' 
current component are more model--dependent than those involving 
only the ``good'' component. In Ref.~\cite{Frankfurt:1981mk} 
this problem was addressed by eliminating 
the ``bad'' ($-$) component using gauge invariance, and applying a trick
to the transverse component. The extension of this technique to the
case of two--photon exchange processes is an interesting problem
but beyond the scope of the present paper.
\section{Numerical estimates}
\label{sec:numerical}
For a numerical estimate of the asymmetry we need to specify the 
spin/flavor wave function of the nucleon. Since Eq.~(\ref{A_N_comp}) 
was derived for the idealized case of a composite nucleon (weak binding), 
the spin--flavor wave function needs to be modeled consistently with
this approximation. We consider two simple models which meet this
requirement.
\begin{itemize}
\item[(a)] $SU(6)$ \textit{spin/flavor wave function.}
The simplest choice of wave function consistent with the
composite nucleon assumption is the wave function of the 
non--relativistic quark model. In this model, the probabilities
$P_{f\sigma}$ for finding a quark in the proton wave function 
with flavor $f = u, d$ and spin projection $\sigma = +, -$
along the direction of the transverse proton spin, are 
\beq
\textstyle
P_{u+} = \frac{5}{9}, \;\; 
P_{u-} = \frac{1}{9}, \;\;
P_{d+} = \frac{1}{9}, \;\;
P_{d-} = \frac{2}{9},
\label{prob_su6}
\eeq
with $\sum_{f\sigma} P_{f\sigma} = 1$,
see Ref.~\cite{Close:1974ux} and references therein. The probabilities 
for the neutron are obtained by interchanging $u \leftrightarrow d$. 
Neglecting the effect of spin on the quark momentum distributions, we obtain
\be
R &=& \frac{e_u^3 (P_{u+} - P_{u-}) + e_d^3 (P_{d+} - P_{d-})}
{e_u^2 (P_{u+} + P_{u-}) + e_d^2 (P_{d+} + P_{d-})}
\\[2ex]
&=& 
\left\{ \begin{array}{rcrl} \frac{11}{27} &=& 0.41 & \text{(proton),} 
\\[2ex] -\frac{2}{9} &=& -0.22 & \text{(neutron) .}
\end{array} \right. 
\label{R_su6}
\ee
The ratio of the neutron to the proton structure factors,
and thus of the corresponding asymmetries, in this
model is
\beq
\frac{R^p}{R^n} \;\; = \;\; 
-\frac{6}{11} \;\; = \;\; -0.55 .
\eeq
\item[(b)] \textit{Transversity $=$ helicity distributions.}
For a weakly bound nucleon we can neglect sea quarks and assume
the valence quark transversity to be equal to the helicity
distributions. It then becomes possible to evaluate the ratio 
(\ref{R_def}) using phenomenological parametrizations for the unpolarized 
and helicity parton densities. With the parametrizations of 
Refs.~\cite{Gluck:1995yr,Gluck:1998xa} we find
that for $Q^2 \sim \text{few GeV}^2$ the ratio is practically 
independent of $Q^2$, and approximately constant in
the region $0.2 < x < 0.5$, with values
\beq
R(x) \;\; \approx \;\;
\left\{ \begin{array}{ll} \phantom{-} 0.35 & \text{(proton),} 
\\[1ex] -0.2 & \text{(neutron) .}
\end{array} \right. 
\eeq
These values are close to the ones obtained with the $SU(6)$
wave function, Eq.~(\ref{R_su6}).
\end{itemize}
%
%
\begin{figure}
\begin{tabular}{c}
\includegraphics[width=7cm]{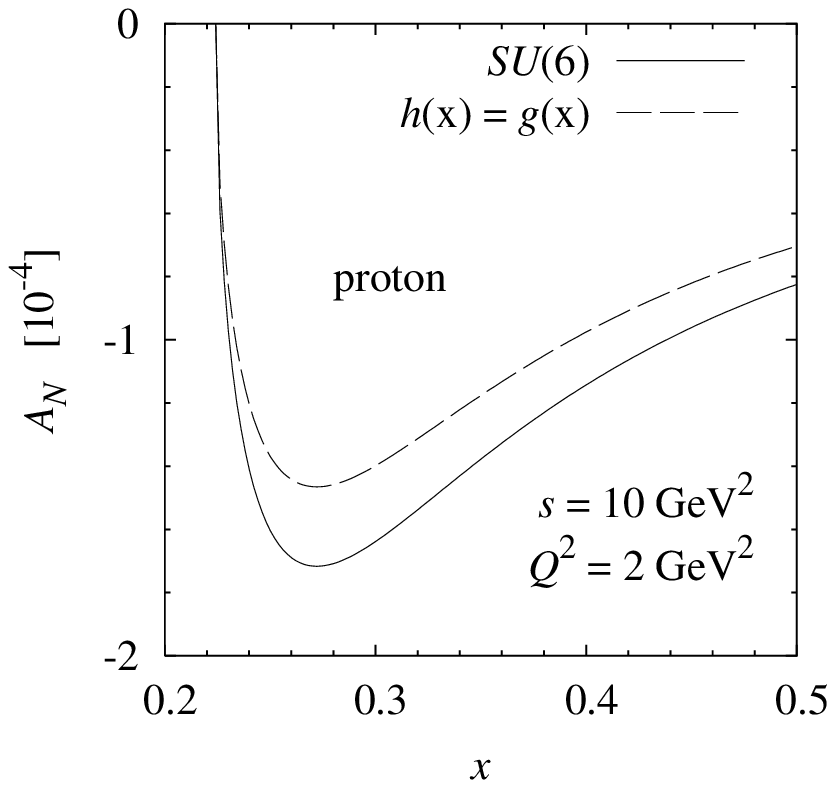}
\\
\includegraphics[width=7cm]{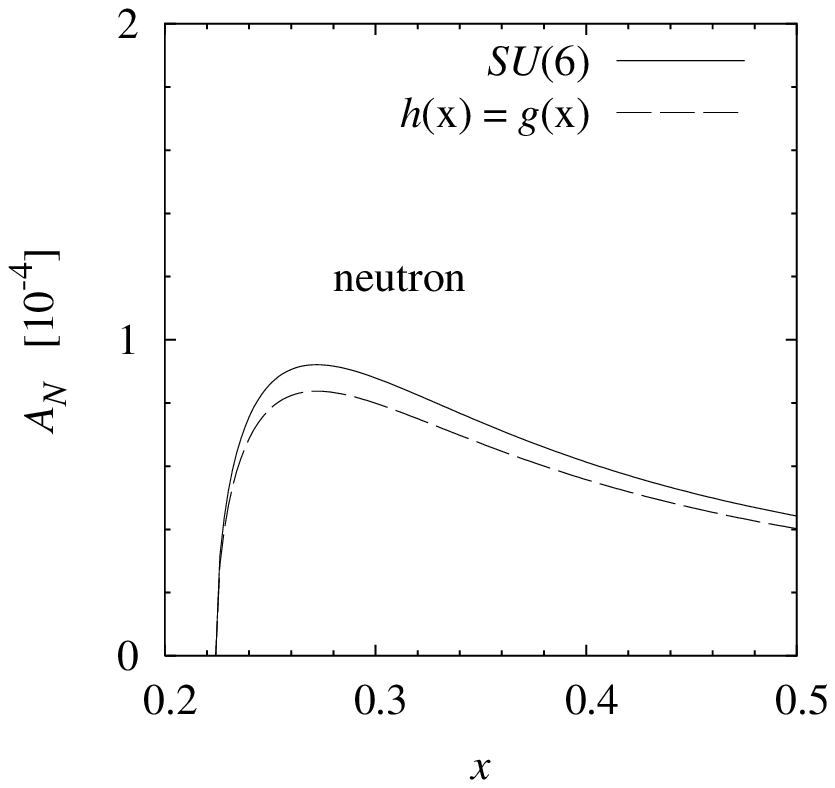}
\end{tabular}
\caption[]{The normal target spin asymmetry $A_N$ 
in DIS kinematics in the composite nucleon approximation, 
Eq.~(\ref{A_N_comp}), for both proton and neutron target,
with different assumptions about the spin--flavor wave function:
(a) $SU(6)$ symmetry, (b) transversity = helicity distributions.
Note the different signs for proton and neutron.
Shown is the asymmetry as a function of $x$, for 
$s = 10 \, \text{GeV}^2$ and $Q^2 = 2 \, \text{GeV}^2$.
The values of $x$ are kinematically restricted to 
$x > x_{\textrm{min}}$, 
Eq.~(\ref{x_min}).} 
\label{fig:comp_pn}
\end{figure}
It is interesting to note that with both models (a) and (b) the ratio of 
the proton and neutron asymmetries in the composite nucleon approximation 
is numerically not far from the ratio of the proton and neutron 
magnetic moments,
\beq
\frac{\mu^p}{\mu^n} \;\; = \;\; -1.46 .
\eeq
This is what one would expect from the simple classical picture 
of the normal spin asymmetry as being due to the scattering from the
magnetic field generated by the target (see Fig.~\ref{fig:xyz}b).

For a numerical estimate of the asymmetry in the composite nucleon
approximation, Eq.~(\ref{A_N_comp}), we use a constituent quark mass
$M_q = M/3$. Fig.~\ref{fig:comp_pn} shows the asymmetry 
for an electron--proton CM energy of $s = 10 \, \text{GeV}^2$
(corresponding approximately to the planned Jefferson Lab Hall A 
experiment \cite{PR-07-013} with $6 \, \textrm{GeV}$ beam energy), 
for $Q^2 = 2 \, \text{GeV}^2$,
as a function of $x$. Note that for given $s$ and $Q^2$ the minimum value 
of $x$ which is kinematically attainable is given by Eq.~(\ref{x_min}).
Comparison of Fig.~\ref{fig:comp_pn} with Fig.~\ref{fig:an_tdep}
shows that the magnitude of the asymmetry for the composite proton
is reduced by a factor of $\sim 4$ compared to the pointlike
proton approximation. This change results from a combination of
various factors: the quark charges and polarizations in the 
structure factor Eq.~(\ref{R_def}), the change of the target mass 
$M \rightarrow M_q$, and the change of the effective CM energy
$s \rightarrow s_{\text{sub}}$ [the latter effect partly compensates
the change in the target mass; note that the pointlike asymmetry
Eq.~(\ref{A_N_simplified}) is proportional to $M/\sqrt{s}$].
Fig.~\ref{fig:comp_pn} shows the results obtained with assumptions 
(a) and (b) about the spin--flavor wave function of the target. 
One sees that the two models give comparable values of 
the asymmetry for both proton and neutron.
%
%
\begin{figure*}
\begin{tabular}{ccc}
\includegraphics[width=7cm]{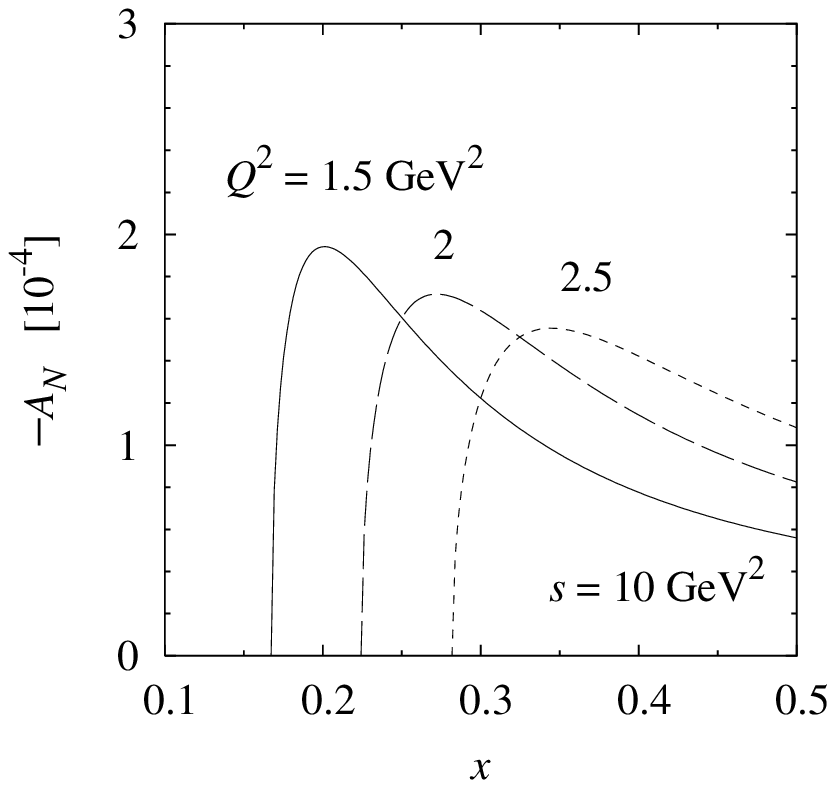}
&
\hspace{1cm}
&
\includegraphics[width=7cm]{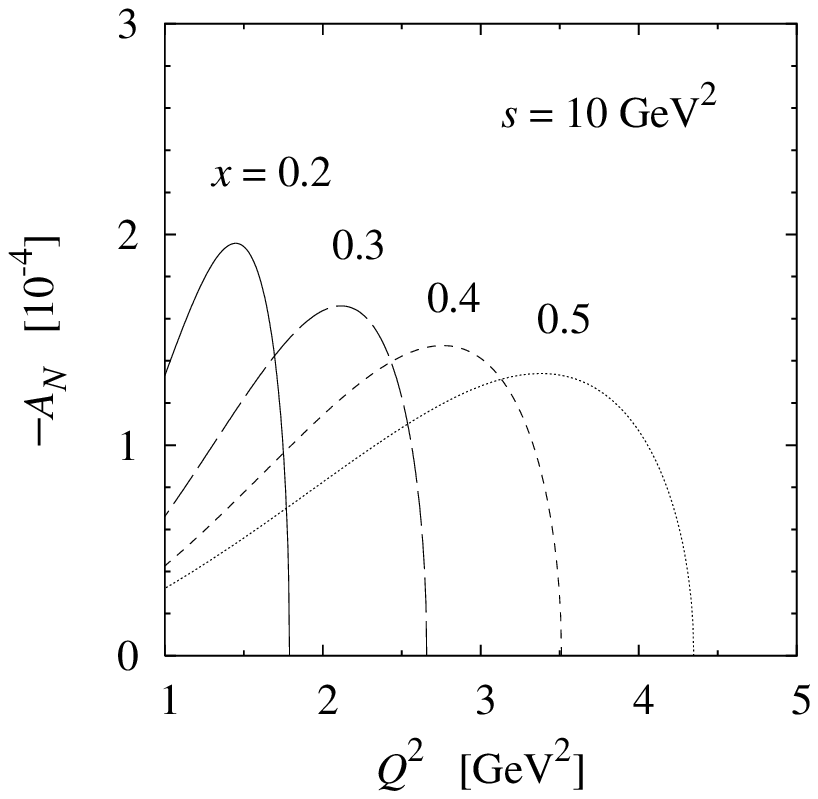}
\end{tabular}
\caption[]{The normal target spin asymmetry $A_N$ 
(note the minus sign on the axis) for the proton 
in the composite nucleon approximation, Eq.~(\ref{A_N_comp}),
as a function of $x$ and $Q^2$. Shown are the results for the 
$SU(6)$ spin/flavor wave function, with 
$s = 10 \, \text{GeV}^2$. \textit{Left:} $x$--dependence 
several values of $Q^2$ (indicated above the curves). 
The values of $x$ are kinematically restricted to 
$x > x_{\textrm{min}}$, Eq.~(\ref{x_min}). \textit{Right:} 
$Q^2$--dependence for several values of $x$ (indicated above the curves).
The values of $Q^2$ are kinematically restricted to 
$Q^2 < Q^2_{\textrm{max}}$, Eq.~(\ref{Q2_max}).} 
\label{fig:comp_xdep_qqdep}
\end{figure*}
%

%
%
\begin{figure}
\includegraphics[width=7cm]{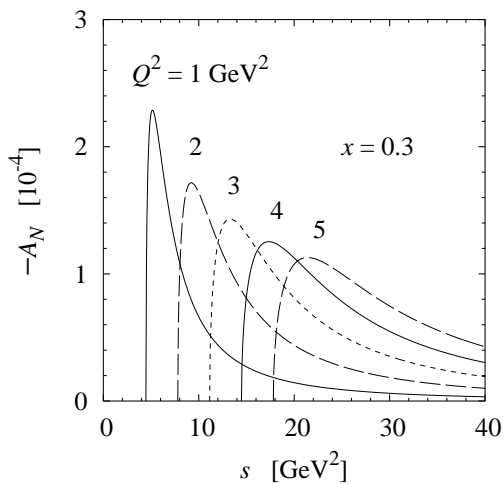}
\caption[]{The normal target spin asymmetry $A_N$ (note the minus sign
on the axis) for the proton in the composite nucleon approximation,
as a function of the squared electron--proton CM energy, 
$s$, for $x = 0.3$ and various values of $Q^2$ (indicated above the curves). 
The asymptotic behavior at large $s$
is $A_N \sim s^{-2}$, with the coefficient proportional to
$Q^3$, \textit{cf.}\ Eqs.~(\ref{A_N_large_s}) and 
(\ref{A_N_large_s_qqdep}).
\label{fig:comp_sdep}}
\end{figure}
Of interest is also the deuteron target. Because of its isoscalar
character, the structure factor (\ref{R_def}) for the deuteron is
\beq
R^d (\xi) \;\; = \;\; \frac{e_u^3 + e_d^3}{e_u^2 + e_d^2}
\; \frac{h^d (\xi)}{f^d (\xi)}
\;\; = \;\; \frac{7}{15} \; \frac{h^d (\xi)}{f^d (\xi)} ,
\eeq
where $h^d \equiv h^d_u = h_d^d$ and $f^d \equiv f^d_u = f_d^d$ 
are the isoscalar quark distributions in the deuteron. Approximating 
the latter by the sum of proton and neutron distributions,
and using isospin invariance, one has
\beq
h^d \;\; \equiv \;\; (h_u + h_d)/2 ,
\hspace{2em}
f^d \;\; \equiv \;\; (f_u + f_d)/2 ,
\eeq
where the distributions without superscript refer to the proton.
With the $SU(6)$ wave functions, \textit{cf.}\ Eq.~(\ref{prob_su6}),
one obtains
\beq
\frac{h^d}{f^d} \;\; = \;\; 
\frac{P_{u+} - P_{u-} + P_{d+} - P_{d-}}
{P_{u+} + P_{u-} + P_{d+} + P_{d-}}
\;\; = \;\; \frac{1}{3} ,
\eeq
and thus
\beq
R^d \;\; = \;\; \frac{7}{45} \;\; \approx \;\; 0.16.
\eeq
The asymmetry for the deuteron has the same sign as for the proton, 
but its magnitude is reduced by a factor $21/55 \approx 0.38$.
With the ``transversity = helicity'' approximation for the
proton and neutron and the parametrizations 
\cite{Gluck:1995yr,Gluck:1998xa} one obtains 
$h^d(\xi) / f^d(\xi) \approx 0.3$ at $\xi = 1/3$, 
very close to the $SU(6)$ result. 

It is interesting to study the dependence on the kinematic variables
of the asymmetry obtained in the composite nucleon, Eq.~(\ref{A_N_comp}).
Figure~\ref{fig:comp_xdep_qqdep} (left panel) shows the asymmetry as a 
function of $x$, for various fixed values of $Q^2$, and fixed $s$. 
One sees that the maximum value of the asymmetry decreases with increasing 
$x$. This is because the magnitude of the asymmetry is inversely proportional
to the invariant CM energy of the quark subprocess, $s_{\text{sub}}^{1/2}$
[\textit{cf.}\ Eq.~(\ref{A_N_simplified}) with 
$s \rightarrow s_{\text{sub}}$], and $s_{\text{sub}}$ is close to $Q^2$ 
at the large subprocess scattering 
angle corresponding to the maximum value of the asymmetry.
Figure~\ref{fig:comp_xdep_qqdep} (right panel) shows the asymmetry
as a function of $Q^2$, for various fixed values of $x$. For fixed
$s$ and $x$, the $Q^2$ range is kinematically restricted to values
lower than Eq.~(\ref{Q2_max}).

Figure~\ref{fig:comp_sdep} shows the dependence of the asymmetry
on the squared electron--nucleon CM energy, $s$, for fixed $x$ and $Q^2$.
This dependence could in principle be tested by comparing measurements at 
different beam energies, similar to the $L/T$ separation of 
electroproduction cross sections. General considerations suggest that
at large $s$ the asymmetry vanishes as $s^{-2}$, \textit{cf}.\ 
Eq.~(\ref{A_N_large_s}) in Sec.~\ref{sec:soft}.
The asymmetry obtained in the composite nucleon approximation
exhibits this behavior, see Fig.~\ref{fig:comp_sdep}.
\section{Transverse spin dependence in soft high--energy scattering}
\label{sec:soft}
In addition to DIS it is interesting to consider the transverse
spin dependence in ``soft'' high--energy scattering, \textit{i.e.},
the limit of large scattering energy, but small energy and momentum 
transfer to the target,
\beq
s \;\; \gg \;\; \mu^2, \hspace{3em} Q^2, \; M_X^2 \sim \mu^2 ,
\label{high_energy}
\eeq
where $\mu$ denotes a typical hadronic mass scale. In this limit
one can analyze the two--photon exchange interference cross section
with general methods for studying the high--energy behavior 
of QED amplitudes. Also in this limit, one can use closure over
non--relativistic quark model states to describe the inclusive final
state and obtain a new perspective on the dominance of single--quark
scattering at larger momentum transfers.

It is well--known that in QED the large--$s$ behavior of 
scattering amplitudes having the form of two blocks connected
by $t$--channel photon exchange is determined only by the number 
of the exchanged photons and their polarization states \cite{Gribov:1970ik}.
The internal structure of the blocks, which themselves do not contain any 
large invariants, is important only insofar it determines which polarization 
states can contribute. One can easily see that for an electron scattering 
from a pointlike target the so-called ``non--sense'' polarization, 
which gives the dominant contribution to the cross section, does not give 
rise to a transverse spin dependence. Indeed, the high--energy 
behavior of the asymmetry for the pointlike target, as follows
from expanding Eq.~(\ref{A_N_simplified}) in the region of
small CM angle, $\theta_{\textrm{cm}} = 2Q/\sqrt{s}$, is
\beq
A_N \;\; = \;\; \frac{\alpha M \, Q^3}{2 s^2}
\hspace{4em} (s \gg Q^2, M^2),
\label{A_N_high_energy}
\eeq
\textit{i.e.}, the asymmetry vanishes in the high--energy limit.
Assuming the polarization structure of the blocks in the pointlike 
target case to be representative of the general case, we conclude that 
the $s$--dependence of the inclusive 
asymmetry should be the same as in the point particle case,
\beq
A_N \;\; \sim \;\; s^{-2}
\hspace{2em} (s \gg \mu^2; \; Q^2, M_X^2 \sim \mu^2 ).
\label{A_N_large_s}
\eeq
A general proof of this statement, adapting the methods of 
Ref.~\cite{Gribov:1970ik}, we leave up to future work.

%
%
\begin{figure}
\includegraphics[width=7cm]{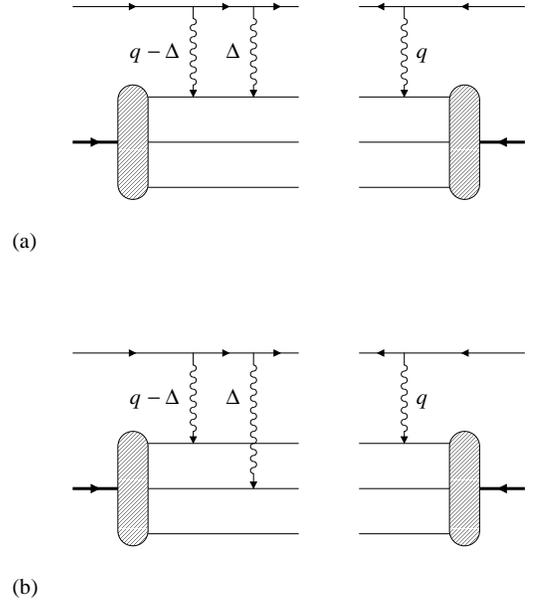}
\caption[]{Interference of one--photon and two--photon exchange in 
electron scattering from a bound state. (a) Two--photon exchange with
the same constituent. (b) Two--photon exchange with different constituents.}
\label{fig:quark}
\end{figure}
The $Q^2$ dependence of the asymmetry in the high--energy limit
(\ref{high_energy}) can be studied using general arguments 
for soft scattering from a composite system based on the mean--field
approximation of nuclear physics, 
see \textit{e.g.}\ Refs.~\cite{Stein:1975yy,Frankfurt:2000jm}.  
In the target rest frame, we consider the nucleon as a generic 
non--relativistic bound state of massive quarks. For sufficiently small
excitation energies one can use closure over the non--relativistic quark
states to calculate the inclusive cross section. In this approach
the electromagnetic coupling of the quark (with label $i$) is described by 
the operator $e_i \, \exp \left[ -i (\bm{q} \bm{r}_i) \right]$, where 
$\bm{q}$ is the photon momentum and $e_i$ and $\bm{r}_i$ the charge and 
position of the quark. Consider now the two contributions to the 
spin--dependent inclusive cross section shown in Fig.~\ref{fig:quark}, 
where the momenta in the two--photon exchange amplitude are denoted by 
$\bm{q} - \bm{\Delta}$ and $\bm{\Delta}$. In the mean--field approximation,
they are proportional, respectively, to 
\be
f^{\text{(a)}}(\bm{\Delta}) &\propto& \sum_{i} e_i^3 , \\
f^{\text{(b)}}(\bm{\Delta}) &\propto& \sum_{i \neq j} e_i^2 e_j \; 
F^2 (\gamma \bm{\Delta}^2) ,
\label{mean_field_a_b}
\ee
where $F$ denotes the elastic form factor of the ground state in 
mean--field approximation, and $\gamma$ is a coefficient of order unity 
which results from the calculation of the recoil of the 
spectator system \cite{Frankfurt:2000jm}.
Note that the form factor appears only in the contribution (b), 
where the two photons couple to different quarks, because here
the momentum $\bm{\Delta}$ has to be routed through the nucleon 
wave function. To get the cross sections, the functions $f^{\text{(a)}}$ 
and $f^{\text{(b)}}$ are to be integrated over $\bm{\Delta}$, together
with the photon propagators and the numerator factors accounting for 
the kinematic momentum dependence of the asymmetry. The IR finiteness
of the spin--dependent cross section now guarantees that the no large
contributions arise from momenta $|\bm{\Delta}| \lesssim R_N^{-1}$
($R_N$ is the size of the bound state), because of the vanishing
of the numerators. Thus, in the region of moderately large momentum 
transfers, $s \gg Q^2 \gg R_N^{-2}$, the relative magnitude of the
contributions (a) and (b) is essentially determined by the phase space 
available for the $\bm{\Delta}$--integral. In case (a) the
integral extends up to $|\bm{\Delta}|^2 \sim Q^2$, while in case (b) 
it is limited to $|\bm{\Delta}|^2 \lesssim R_N^{-2}$ by the form factors.
We conclude that in the region $s \gg Q^2 \gg R_N^{-2}$ the dominant
contribution to the spin--dependent cross section comes from the
coupling of the photons to the same quark; the contribution in which
the two photons couple to different quarks is suppressed by a
high power of $1/(R_N^2 Q^2)$, which depends on the detailed behavior 
of the bound--state form factor. Similar arguments apply to the other 
possible contributions to the interference cross section 
(not shown in Fig.~\ref{fig:quark}) in which not all photons 
couple to the same quark.

The spin--independent cross section in one--photon exchange approximation
is likewise dominated by the scattering of the two photons from the
same quark; interference contributions are suppressed by the bound--state
form factor. In this case the above reasoning just reproduces standard 
arguments for the approach to scaling in the non--relativistic quark model. 
Combining the statements about the spin--dependent and spin--independent
cross sections for a composite target, we conclude that the asymmetry 
(\textit{i.e.}, the ratio) should exhibit the same $Q^2$--dependence as 
the asymmetry for a point particle, Eq.~(\ref{A_N_high_energy}),
\beq
A_N \;\; \sim \;\; Q^3
\hspace{2em} (s \gg Q^2 \gg R_N^{-2}).
\label{A_N_large_s_qqdep}
\eeq
In summary, our arguments based on the mean--field approximation 
imply that the $Q^2$--dependence of the asymmetry at moderately large 
$Q^2$ is the minimal dependence dictated by kinematics (\textit{i.e.}, 
by the need to have transverse momentum transfer) but is not subject 
to any dynamical form factor suppression. We note that the asymmetry 
calculated in the constituent quark model with the composite nucleon 
approximation (see Sec.~\ref{sec:composite}) shows the behavior described 
by Eqs.~(\ref{A_N_large_s}) and (\ref{A_N_large_s_qqdep}).
\section{Summary and outlook}
\label{sec:summary}
The transverse spin dependence of the cross section of inclusive 
$eN$ scattering, in spite of being a ``simple'' observable, is seen
to give rise to many interesting questions of electrodynamics and
strong interaction physics. Our treatment of these problems in large 
parts has been of exploratory nature. Following we summarize our main
conclusions, and describe several problems deserving further study.

Concerning the electrodynamics aspects, we have pointed out that the
transverse spin--dependent cross section due to two--photon exchange 
is free of IR divergences. No cancellation of IR divergences between 
two--photon exchange and real photon emission is required 
(as in the two--photon corrections to the spin--independent cross section), 
making the transverse spin--dependent cross section a clean 
two--photon exchange observable. However, real photon emission
can still make a finite contribution to the spin dependence of
$ep \rightarrow e'X$, which in practice cannot be separated from 
purely hadronic final states. To estimate this contribution
is an interesting problem for further study.

Concerning the strong interaction aspects, we have argued that in DIS
kinematics a sizable contribution to the transverse spin--dependent 
cross section results from quark helicity--flip processes
made possible by the non--perturbative vacuum structure of QCD
structure (chiral symmetry breaking). The key point is that
such processes are not significantly Sudakov--suppressed if the IR cutoff 
for gluon emission is of the order of the chiral symmetry 
breaking scale $\mu^2_{\textrm{chiral}} \gg \Lambda^2_{\textrm{QCD}}$.
While this seems natural in the context of the phenomenology of chiral
symmetry breaking, we presently cannot offer rigorous arguments for
the correctness of this choice.

We have presented qualitative arguments why the quark helicity--conserving 
contribution to the transverse spin--dependent cross section, 
related to $g_{T, f}$, is unlikely to dominate. A complete QCD calculation 
of this contribution in the collinear factorization approach, 
which maintains electromagnetic gauge invariance by including 
quark--gluon operators and avoids unphysical collinear divergences, 
is clearly an outstanding problem. The crucial question is whether the 
complete result will involve the Wandzura--Wilczek contribution to 
$g_{T, f}$ (which is given in terms of matrix elements of twist--2 
operators), or whether only the twist--3 quark--gluon correlations 
will survive. In the former case the helicity--conserving contribution 
could be estimated in a model--independent way. In the latter case, 
it is likely to be very small, and the dominant contribution to the 
transverse spin--dependent cross section would most likely come from 
the quark helicity--flip process governed by the transversity distribution. 

Our numerical estimates based on the constituent quark model
suggest that the asymmetry in the kinematics of the planned Jefferson Lab
Hall A experiment \cite{PR-07-013} is of the order a few times $10^{-4}$, 
with different sign for proton and neutron. The predicted asymmetry 
for the proton is larger than for the neutron, suggesting that 
measurements with a transversely polarized proton target would 
be a useful complement to the planned measurements with $^{3}\textrm{He}$.
\begin{acknowledgments}
The authors thank X.~Jiang for initiating this investigation,
and T.~Holmstrom, A.~Metz, and M.~Schlegel for useful discussions.
This work is supported by DOE. A.~A.\ acknowledges partial support 
by NSF under grants PHY--0114343 and PHY--0301841.

Notice: Authored by Jefferson Science Associates, LLC under U.S.\ DOE
Contract No.~DE-AC05-06OR23177. The U.S.\ Government retains a
non--exclusive, paid--up, irrevocable, world--wide license to publish or
reproduce this manuscript for U.S.\ Government purposes.
\end{acknowledgments}

\begin{thebibliography}{99}
\bibitem{foot:absorptive} The precise definition of the absorptive part 
of the hadronic Compton amplitude for the $N \rightarrow X$ transition
in a non--perturbative context is not needed for the following. Later we 
shall evaluate Eq.~(\ref{Im_M_integral}) for a field--theoretical model
of hadron structure, in which the absorptive part is given explicitly 
as the sum of Cutkosky cuts of Feynman diagrams.
\bibitem{foot:ftmodel} 
Since we are interested only in the absorptive part of the two--photon
exchange amplitude, ``seagull'' terms with two photons coupling to the 
scalar at the same point do not contribute.
\bibitem{foot:mass}
It is worthwhile pointing out that our arguments for the role of 
non--perturbative helicity flip in the transverse spin dependence are not 
in conflict with the experience with QCD factorization for the usual DIS 
cross sections, where no non--perturbative quark mass is included. 
In the unpolarized cross sections ($F_2, F_L$) the dominant contribution 
comes from the helicity--conserving leading--twist contribution, and a 
non--perturbative quark mass would result only in a 
higher--twist correction of the order $M_q^2/Q^2$.
\end{thebibliography}
\end{document}